\documentclass[Numbered,pdflatex,sn-nature]{sn-jnl}

\usepackage{graphicx}%
\usepackage{multirow}%
\usepackage{amsmath,amssymb,amsfonts}%
\usepackage{amsthm}%
\usepackage{mathrsfs}%
\usepackage[title]{appendix}%
\usepackage{xcolor}%
\usepackage{textcomp}%
\usepackage{manyfoot}%
\usepackage{booktabs}%
\usepackage{algorithm}%
\usepackage{algorithmicx}%
\usepackage{algpseudocode}%
\usepackage{listings}%
\usepackage{caption}

\theoremstyle{thmstyleone}%
\theoremstyle{thmstyletwo}%

\theoremstyle{thmstylethree}%

\usepackage{CJKutf8}
\usepackage{amsmath}

\newcommand{\vesc}{v$_{esc}$}

\newcommand{\oiiitext}{[O~{\sc iii}]}

\newcommand{\lya}{Ly$\alpha$}

\newcommand{\Lbol}{$L_{\rm bol}$}
\newcommand{\edd}{$\lambda_{\rm Edd}$}
\newcommand{\mbh}{$M_{\rm BH}$}
\newcommand{\arcsec}{$^{\prime\prime}$}

\newcommand{\mum}{\ifmmode{\rm \mu m}\else{$\mu$m}\fi}
\newcommand{\vdisp}{$\vdisp$}

\newcommand{\wjiu}{w$_{90}$}
\newcommand{\vwu}{{v$_{50}$}}
\newcommand{\vjiu}{{v$_{90}$}}

\newcommand{\Lwu}{{$\lambda L_{\lambda}(5100)$}}

\newcommand{\vjiuba}{{v$_{98}$}}
\newcommand{\vlingwu}{{v$_{05}$}}
\newcommand{\vjiuwu}{{v$_{95}$}}

\newcommand{\oiii}{{[O~{\sc iii}] $\lambda$5007}}
\newcommand{\oiiiab}{{[O~{\sc iii}] $\lambda$$\lambda$4959,5007}}

\newcommand{\sii}{\hbox{[S$\,${\scriptsize II}]}}
\newcommand{\siitext}{\hbox{[S$\,${\scriptsize II}]}}

\newcommand{\hb}{\hbox{H$\beta$}}
\newcommand{\hg}{\hbox{H$\gamma$}}

\newcommand{\kms}{km s$^{-1}$} 
\newcommand{\msun}{M$_{\odot}$} 
\newcommand{\msunyr}{{M$_{\odot}$ yr$^{-1}$}}

\begin{document}

\title{Extreme Galaxy-scale Outflows Are Frequent among Luminous Early Quasars}

\author*[1]{\fnm{Weizhe} \sur{Liu}}\email{oscarlwz@gmail.com}
\affil[1]{\orgdiv{Steward Observatory}, \orgname{University of Arizona}, \orgaddress{\street{933 N Cherry Ave}, \city{Tucson}, \postcode{85721}, \state{AZ}, \country{USA}}}

\author[1]{\fnm{Xiaohui} \sur{Fan}}

\author[2]{\fnm{Huan} \sur{Li}}
\affil[2]{\orgdiv{School of Aerospace Science and Technology}, \orgname{Xidian University},
\orgaddress{\city{Xian}, \postcode{710126}, \country{China}}}

\author[1]{\fnm{Richard} \sur{Green}}

\author[3]{\fnm{Jinyi} \sur{Yang}}
\affil[3]{\orgdiv{Department of Astronomy}, \orgname{University of Michigan}, \orgaddress{\street{1085 S. University, 323 West Hall}, \city{Ann Arbor}, \postcode{48109-1107}, \state{MI}, \country{USA}}}

\author[1,3]{\fnm{Xiangyu} \sur{Jin}}

\author[1]{\fnm{Jianwei} \sur{Lyu}}

\author[1]{\fnm{Maria} \sur{Pudoka}}

\author[1]{\fnm{Yongda} \sur{Zhu}}


\author[4]{\fnm{Eduardo} \sur{Ba\~nados}}
\author[4,7]{\fnm{Silvia} \sur{
Belladitta}}
\affil[4]{\orgname{Max Planck Institut f\"ur Astronomie}, \orgaddress{K\"onigstuhl 17}, \postcode{D-69117} \city{Heidelberg}, \country{Germany}}

\author[5]{\fnm{Thomas} \sur{Connor}}
\affil[5]{\orgname{Center for Astrophysics $\vert$\ Harvard\ \&\ Smithsonian}, \orgaddress{\street{ 60 Garden St.}, \city{Cambridge}, \postcode{02138}, \state{MA}, \country{USA}}}

\author[6]{\fnm{Tiago} \sur{Costa}}
\affil[6]{\orgdiv{School of Mathematics, Statistics and Physics},  \orgname{Newcastle University}, \orgaddress{\street{NE1 7RU}, \country{UK}}}

\author[7]{\fnm{Roberto} \sur{Decarli}}
\affil[7]{\  \orgname{
INAF – Osservatorio di Astrofisica e Scienza dello Spazio
di Bologna}, \orgaddress{\street{via Gobetti 93/3, I-40129}, \city{Bologna}, \country{Italy}}}

\author[8,9]{\fnm{
Anna-Christina} \sur{Eilers}}
\affil[8]{\orgdiv{Department of Physics}, \orgname{Massachusetts Institute of Technology}, \orgaddress{\city{Cambridge}, \state{MA} \postcode{02139}, \country{USA}}}
\affil[9]{\orgdiv{MIT Kavli Institute for Astrophysics and Space Research}, \orgname{Massachusetts Institute of Technology}, \orgaddress{\city{Cambridge}, \state{MA} \postcode{02139}, \country{USA}}}

\author[10,11]{\fnm{Hyunsung D.} \sur{Jun}}
\affil[10]{\orgdiv{Department of Physics}, \orgname{Northwestern College}
\orgaddress{\street{101 7th Street SW}, \city{Orange City}, \state{IA}, \postcode{51041}, \country{USA}}}
\affil[11]{\orgdiv{School of Physics}, \orgname{Korea Institute for Advanced Study}
\orgaddress{\street{85 Hoegiro, Dongdaemun-gu}, \city{Seoul}, \state{}, \postcode{02455}, \country{Republic of Korea}}}

\author[12]{\fnm{Madeline A.} \sur{Marshall}}
\affil[12]{\orgname{Los Alamos National Laboratory},
\orgaddress{\city{Los Alamos}, \state{NM}, \postcode{87545} \country{USA}}}

\author[13]{\fnm{Chiara} \sur{Mazzucchelli}}
\affil[13]{\orgdiv{Instituto de Estudios Astrof\'{\i}sicos, Facultad de Ingenier\'{\i}a y Ciencias} \orgname{Universidad Diego Portales},
\orgaddress{\street{Avenida Ejercito Libertador 441} \city{Santiago}, \country{Chile}}}

\author[14]{\fnm{Jan-Torge} \sur{Schindler}}
\affil[14]{\orgdiv{Hamburger Sternwarte}, \orgname{University of Hamburg}, \orgaddress{\street{Gojenbergsweg 112}, \city{Hamburg}, \postcode{D-21029}, \country{Germany}}}

\author[15]{\fnm{Yue} \sur{Shen}}
\affil[15]{\orgdiv{Department of Astronomy}, \orgname{University of Illinois at Urbana-Champaign}, \orgaddress{\city{Urbana}, \state{IL} \postcode{61801}, \country{USA}}}

\author[16]{\fnm{Sylvain} \sur{Veilleux}}
\affil[16]{\orgdiv{Department of Astronomy and Joint Space-Science Institute}, \orgname{University of Maryland}, \orgaddress{\city{College Park}, \state{MD}, \postcode{20742}, \country{USA}}}

\author[4]{\fnm{Julien} \sur{Wolf}}

\author[17]{\fnm{Huanian} \sur{Zhang}}
\affil[17]{\orgdiv{Department of Astronomy}, \orgname{Huazhong University of Science and Technology}, \orgaddress{ \city{Wuhan}, \postcode{430074}, \state{Hubei}, \country{China}}}

\author[15]{\fnm{Mingyang} \sur{Zhuang}}

\author[18]{\fnm{Siwei} \sur{Zou}}
\affil[18]{\orgdiv{Chinese Academy of Sciences South America Center
for Astronomy, National Astronomical Observatories}, \orgname{CAS}, \orgaddress{\city{Beijing}, \postcode{100101}, \country{China}}}

\author[19]{\fnm{Mingyu} \sur{Li}}
\affil[19]{\orgdiv{Department of Astronomy}, \orgname{Tsinghua University},
\orgaddress{ \city{Beijing}, \postcode{100084}, \country{China}}}

\maketitle

\textbf{The existence of abundant post-starburst/quiescent galaxies just $\sim$1--2 Gyrs after the Big Bang challenges our current paradigm of galaxy evolution \cite{Carnall2023,Glazebrook2024,deGraff2025}. Cosmological simulations suggest that quasar feedback is likely the most promising mechanism responsible for such rapid quenching \cite{Dubois2013b,Hartley2023,Lovell2023}. Here we report a high detection rate (6/27) of exceptionally fast and powerful galaxy-scale outflows traced by \oiiitext\ emission in z $\sim$ 5--6 luminous quasars as revealed by the James Webb Space Telescope (JWST), with velocity up to $\sim$8400 \kms\ and order-of-magnitude kinetic energy outflow rates up to $\sim$260\% the observed quasar bolometric luminosities.
{This fraction is $>$3.9 and $\sim$8.8 times of those in comparison samples at z $\sim$ 1.5--3.5 and z $<$ 1, respectively.}
These extreme outflows are comparable to or even faster than the most rapid \oiiitext\ outflows reported at z $\lesssim$ 3, and could reach the circumgalactic medium (CGM) or even the intergalactic medium (IGM). The average kinetic energy outflow rate of our sample is {more than 2 dex higher than those of the lower-redshift comparison samples}. The substantially higher frequency of outflows with energetics well above the threshold for negative feedback in our sample strongly suggests that quasar feedback plays a significant role in efficiently quenching/regulating early massive galaxies.}

Luminous quasars already exist just $\lesssim$1 Gyr after the Big Bang \cite{Fan2023} and could heavily impact the systems they reside in. Intriguingly, recent JWST observations have revealed post-starburst activity in two luminous quasars at $z\sim6$, hinting at quasar feedback-induced quenching at early epoch \cite{Onoue2024}.   
Frequent quasar-driven outflows, as a critical form of quasar feedback, are predicted by {simulations} at $z\gtrsim5$ \cite{Costa2018b,Lupi2022}.
{In the pre-JWST era, observations of such outflows
can be mainly divided into two categories: some studies focus on the nuclear winds traced by rest-frame ultraviolet (UV) broad absorption lines (BAL) \cite{Bischetti2022} and emission lines \cite{Shen2019,Yang2021}, with an enhanced BAL incidence rate reported at z$\sim$6 \cite{Bischetti2022}. While fast, the radial distances of BAL and thus their immediate impact on the galaxy scale are not constrained at this epoch, which may be trivial if they are on pc-scale as seen in many lower-redshift cases \cite{Capellupo2011}.
Nevertheless, recent observations imply tantalizing evidence of a connection 
between BAL and large-scale outflows at $z\sim5.5$ \cite{Zhu2025}. }
Other studies utilize rest-frame far-infrared (FIR) emission and absorption lines \cite{Spilker2025}. 
They are confined to FIR-bright objects and inconsistent results on the presence of outflow are reported for both the most promising individual cases \cite{Maiolino2012,Meyer2022} and in stacking analysis \cite{Bischetti2019,Novak2020}. 
With the advent of JWST, \oiii\ emission lines in the rest-frame optical, widely adopted as an unambiguous tracer of galaxy-scale (kpc-scale) quasar/{Active Galactic Nucleus (AGN)} outflows at $z\lesssim3$ \cite{zaka16b,Liu2020,VeilleuxLiu2023}, are finally accessible for objects at $z\gtrsim5$ \cite{Marshall2023,Yang2023b,Loiacono2024,Decarli2024,Liu2024b,Lyu2025}. 
As forbidden transitions, \oiiitext\ emission lines are located on scales $\gtrsim$1 kpc in luminous quasars like our objects {(see \textit{Outflow Properties} in \textit{Methods} for more details)}. This spatial scale is comparable to the host galaxy sizes ($\sim$2 kpc) of early quasars in the rest-frame optical \cite{yue_eiger_2024}.

\begin{figure*}[ht]
\centering
\includegraphics[width=0.9\textwidth]{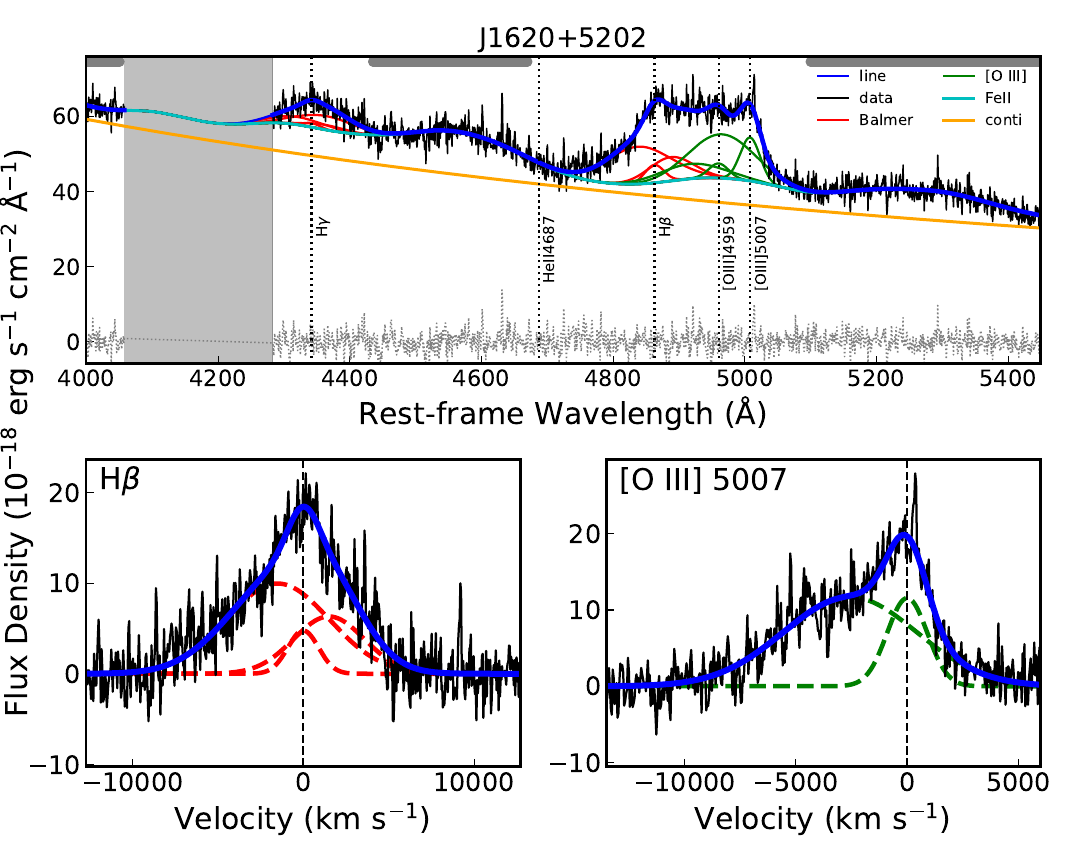}
\caption{\textbf{Spectrum of J1620$+$5202.} \textit{Top:} The object with the fastest \oiiitext\ outflow ($|$\vjiuba$|$ $\sim$ 8400 \kms) discovered in our sample, with the rest-frame JWST spectrum (black), best-fit emission line profiles (blue), iron emission (cyan), continuum (orange), and residual (gray dotted line). The best-fit individual Gaussian components for \hb\ and \hg\ are shown in red and those for \oiiiab\ are shown in green. Systemic velocities of individual emission lines are shown in vertical black dotted lines. The spectral windows adopted for fitting the quasar pseudo continuum are marked by the gray thick bars. The detector gap and adjacent noisy regions not used in the fitting are masked by the vertical gray shaded region. \textit{Bottom:} \hb\ (left) and \oiii\ (right) line profiles with their best-fit models (blue solid lines) and individual components (dashed lines).}
\label{fig:profile}
\end{figure*}

\begin{figure*}[!h]
\begin{minipage}[t]{0.5\textwidth}
 \centering
\includegraphics[width=\textwidth]{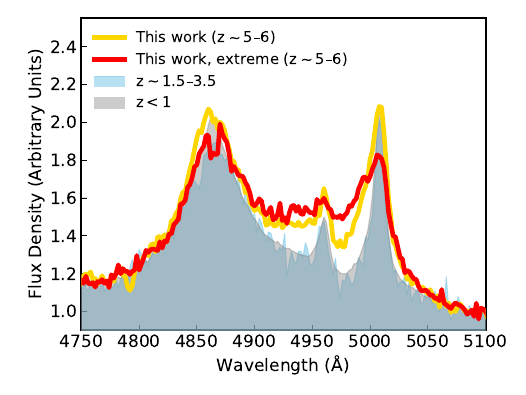}
\end{minipage}
 \begin{minipage}[t]{0.5\textwidth}
 \centering
\includegraphics[width=\textwidth]{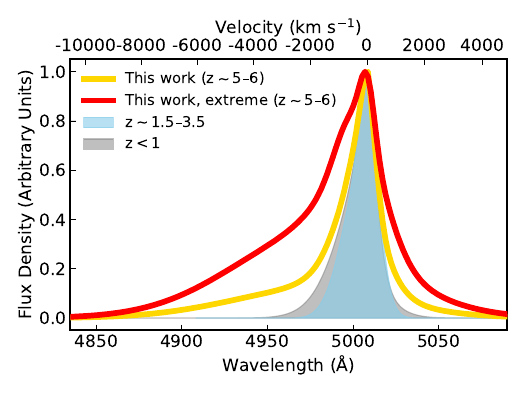}
\end{minipage}

 \begin{minipage}[t]{0.5\textwidth}
 \centering
\includegraphics[width=\textwidth]{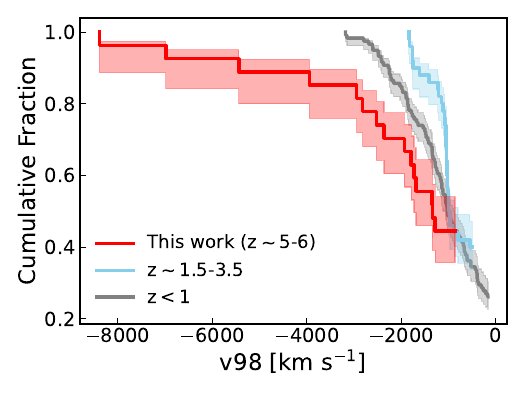}
\end{minipage}
 \begin{minipage}[t]{0.5\textwidth}
 \centering
\includegraphics[width=\textwidth]{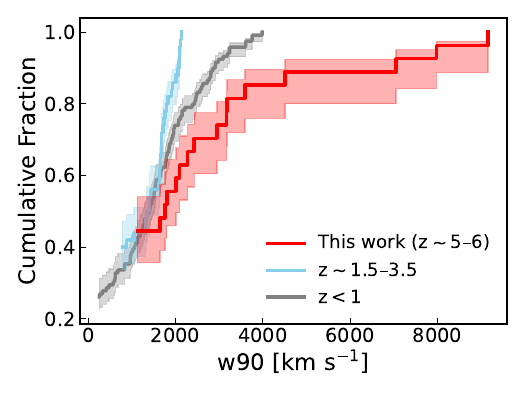}
\end{minipage}

\caption{\textbf{Comparison of emission line profiles and outflow kinematics.} \textit{Upper Left:} 
Mean rest-frame spectra of outflows (orange) and extreme outflows ($|$\vjiuba$|$$>$2700 \kms; red) in our sample, zoomed in to the \hb--\oiiitext\ region, in comparison with outflow sources {in the \textit{z $\sim$ 1.5--3.5 sample} (blue) and \textit{z $<$ 1 sample} (gray).} The spectra of individual objects are binned to 2 \AA/pixel. The mean spectra are normalized with the mean flux density within 5080--5100\AA. \textit{Upper Right:} Same as left but for the best-fit \oiii\ model profile, which are instead normalized at maximum flux density.
\textit{Bottom:} 
Cumulative distribution functions of \oiiitext\ non-parametric kinematics measurements (left: \vjiuba; right: \wjiu) for our sample (red), {the \textit{z $\sim$ 1.5--3.5 sample} (blue) and the \textit{z $<$ 1 sample} (gray)}. The 68\% confidence intervals are indicated by the shaded regions.}  
\label{fig:hist}
\end{figure*}

\begin{figure*}
 \begin{minipage}[t]{0.49\textwidth}
 \centering
\includegraphics[width=\textwidth]{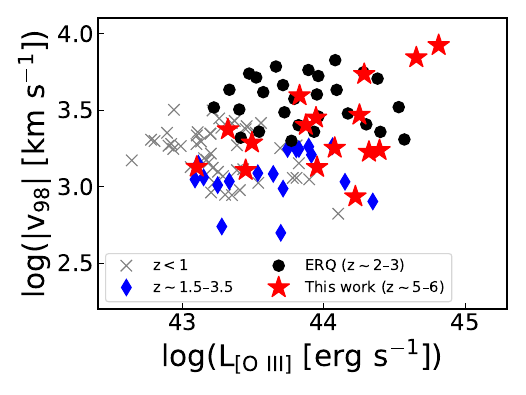}
\end{minipage}
\begin{minipage}[t]{0.49\textwidth}
 \centering
\includegraphics[width=\textwidth]{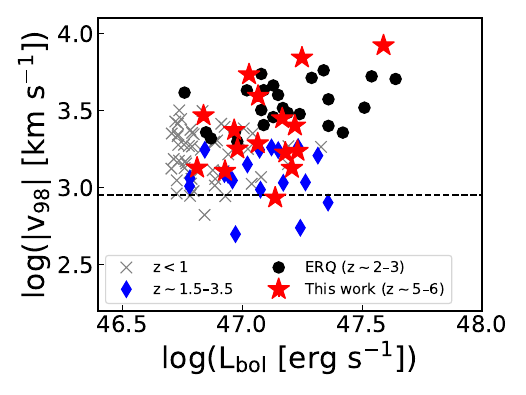}
\end{minipage}
\caption{\textbf{Outflow velocity as a function of quasar luminosity.} $|$\vjiuba$|$ as a function of \oiiitext\ luminosity (\textit{left}) and \Lbol\ (\textit{right}) for our sample (red stars), the ERQs at z $\sim$ 2--3 (black), the {\textit{z $\sim$ 1.5--3.5 sample} (blue) and the z $<$ 1 sample (gray). 
The typical uncertainties of \vjiuba\ are smaller than the symbol sizes.} 
The black dashed line indicates the estimated maximum escape velocity for these quasars.} 
\label{fig:outflow}
\end{figure*}

We have conducted a shallow JWST/NIRSpec {integral field unit (IFU)} survey of 27 luminous Type 1 quasars at z $\sim$ 5--6 to probe the rest-frame optical quasar spectral properties  ({Q-IFU}; see \textit{Observation and Data Reduction} in \textit{Methods}). Our data have revealed blueshifted and broad \oiiitext\ emission lines in 16 objects. 
{In this study, we broadly classify all such objects with 50th percentile velocity \vwu\ $<$ 0 \kms\ as outflows (see \textit{Velocity Thresholds for Outflows and Extreme Outflows} in \textit{Methods} for more discussions).} 
The most extreme example of these outflows is shown in Fig. \ref{fig:profile}. 
{We compare their outflow properties with two representative luminous Type 1 quasar samples at lower-redshift with \oiiitext\ spectroscopic data of similar quality and falling within the same bolometric luminosity range (log$(L_{\rm bol}$/erg s$^{-1}$) $=$ 46.7--47.7), BH mass range (log(M$_{\rm BH}$/\msun) $=$ 8.6--9.7) and Eddington ratio range (\edd\ $=$ 0.1--2.6) as our sample: The \textit{z $\sim$ 1.5--3.5 sample} with 50 objects from \cite{Shen2016} and the \textit{z $<$ 1 sample} with  119 objects from \cite{Wu2022} (See \textit{Comparison Samples} in \textit{Methods} for more details).}
On average, the \oiiitext\ line of outflow objects in our high-redshift sample is significantly more blueshifted and broader than those in the two lower-redshift samples (top panels of Fig. \ref{fig:hist}). 
We adopt a non-parametric approach to describe the gas kinematics (see \textit{Outflow Properties} in \textit{Methods}) and define extreme outflows as those with velocity $|$\vjiuba$|$$>$2700 \kms, exceeding 3$\times$ escape velocity at 1 kpc of a 10$^{13}$ \msun\ dark matter halo {(See \textit{Velocity Thresholds for Outflows and Extreme Outflows} in \textit{Methods} for more details)}.
Such extreme outflows occur in 6 out of 27 in our sample at z$\sim$5--6 or 22.2$_{-9.9}^{+15.7}$\%  (The spectra of the remaining 5 extreme outflows can be found in Extended Data Fig. \ref{fig:samples}). {This fraction is $>$3.9$\times$ ($>$2.2$\times$ -- $>$6.7$\times$) of that in the \textit{z $\sim$ 1.5--3.5 sample}: 0/50 or $<$5.7\%, and $\sim$8.8$\times$ (2.0$\times$ -- 33.0$\times$) of that in the \textit{z $<$ 1 sample}: 3/119 or 2.5$_{-1.4}^{+3.8}$\%.} Here the uncertainties and upper limits are at 90\% confidence level and derived adopting a binomial distribution with Bayesian approach \cite{Cameron2011}. The \oiiitext\ detection rates in the three quasar samples are comparable, so the higher detection rate of extreme outflows in our sample is not resulted from a higher \oiiitext\ detection rate. Moreover, these extreme outflows are generally extended on kpc scale (see \textit{Outflow Properties} in \textit{Methods}), implying that their influence is on galaxy scale.
In the bottom panels of Fig. \ref{fig:hist}, we compare the cumulative distributions of \oiiitext\ kinematics of our objects with the other two samples, which again shows a clear excess of high-velocity outflows in our high-z sample. 
This is confirmed by the Mann–Whitney \textit{U} tests (with the null hypothesis that the two samples are drawn from the same distribution; \cite{Utest}) suggesting that the \vjiuba\ and \wjiu\ distributions of our sample are different from the other two samples, {with $p$-values of $\sim$2.8$\times$10$^{-5}$ and $\sim$1.3$\times10^{-4}$ for comparisons with the \textit{z $\sim$ 1.5--3.5 sample}, and $\sim$3.9$\times$10$^{-4}$ and $\sim$1.1$\times$10$^{-3}$ for comparisons with the \textit{z $<$ 1 sample}, respectively.}

At lower redshifts, such extreme outflows are only found in extraordinary objects like extremely red quasars (ERQs), which exhibit among the fastest/most powerful outflows known at/before cosmic noon, providing compelling evidence of quasar feedback shaping the evolution of massive galaxies \cite{zaka16b,Perrotta2019,Vayner2023c}. ERQs are a population of luminous dusty quasars defined by their extremely red colors in the rest-frame UV to mid-IR, i$-$W3 $>$ 4.6 \cite{Ross12,Hamann2017}. They are rare, with only 205/173636 ($\sim$0.12\%) quasars at 2 $<$ z $<$ 3.4 from the SDSS \citep{Hamann2017}. While they share similar \Lbol\, \mbh, and \edd\ to our sample \cite{Perrotta2019}, their dusty nature makes them distinctly different from blue, unabsorbed Type 1 quasars in our sample and the other two lower-redshift samples. To better understand the properties and impact of outflows in our objects, in Fig. \ref{fig:outflow}, we further compare the {outflow velocities} of our sample with those among ERQs {falling within the same \Lbol\ range as our sample, along with the other two lower-redshift samples.}
Looking at the 4 samples altogether, their {outflow velocities} show large scatters at similar \oiiitext\ and bolometric luminosities. The results are similar for \mbh\ and Eddington ratios. Looking at our sample alone, the correlations between {outflow velocities} and the aforementioned quasar properties are still weak ($p$-values $\sim$ 0.1--0.6 from the Kendall's tau test \cite{Kendall} with the null hypothesis that the two variables are uncorrelated) over the limited dynamical range. 
The median outflow velocity of our objects ($|$\vjiuba$|$ $\sim$ 2150 \kms) is lower than that of ERQs ($|$\vjiuba$|$ $\sim$ 3300 \kms) but larger than those found in the \textit{z $\sim$ 1.5--3.5 sample} ($|$\vjiuba$|$ $\sim$ 1050 \kms) and \textit{z $<$ 1 sample} ($|$\vjiuba$|$ $\sim$ 1280 \kms). Four objects from our sample have $|$\vjiuba$|$ larger than the median value of ERQs, and the most extreme outflow in our sample ($|$\vjiuba$|$ $\sim$ 8400 \kms) is faster than the maximum value reported in the ERQs ($|$\vjiuba$|$ $\sim$ 6700 \kms). In short, compared with similar lower-redshift, blue Type 1 quasars with comparable luminosities, z$\sim$5--6 quasars (as a population) show a significantly higher frequency of extreme outflows with velocities among the highest reported so far, implying more substantial impact to their host galaxies.

The early galaxies are in general compact with sizes of only a few kpc, and thus the outflows in our sample could in principle easily affect a significant portion of their host galaxies and reach to even larger scales. 
As shown in Fig. \ref{fig:outflow}, the {outflow velocities} ($|$\vjiuba$|\sim$2800--8400 \kms) of the 6 extreme objects are significantly faster than the estimated maximum escape velocity ($\sim$600--900 \kms) at 1 kpc of the systems (See \textit{Nuclear Properties and Escape Velocities} in \textit{Methods}).
Therefore, it is expected that a significant portion of the outflowing gas in these 6 objects will reach the CGM and/or IGM, injecting energy, depositing metals, and helping with the creation of the giant \lya-nebulae \cite{farina_requiem_2019,Costa2022}. These outflows may thus play an important role in preventing gas in the CGM from cooling, starving the galaxies and shaping their gaseous environments at z$\gtrsim$5. 

\begin{figure*}[!htb]
\begin{minipage}[t]{0.5\textwidth}
 \centering
\includegraphics[width=\textwidth]{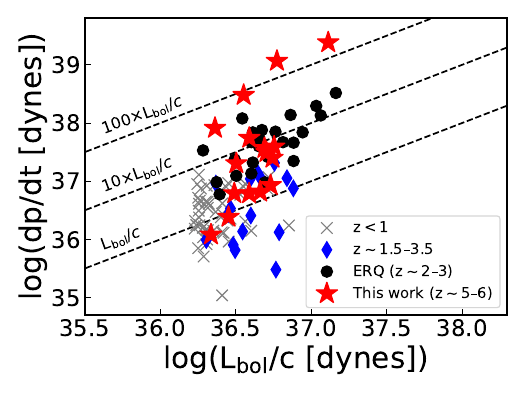}
\end{minipage}
 \begin{minipage}[t]{0.5\textwidth}
 \centering
\includegraphics[width=\textwidth]{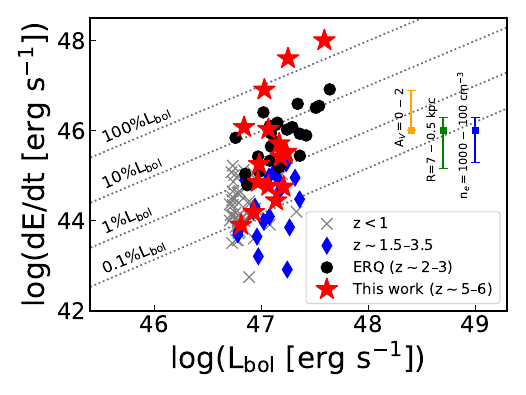}
\end{minipage}
\caption{\textbf{Comparison of outflow energetics.} Momentum outflow rates as a function of quasar radiation force (\textbf{left}) and kinetic energy outflow rates as a function of \Lbol\ (\textbf{right}) for our sample (red), ERQs (black), the {\textit{z $\sim$ 1.5--3.5 sample} (blue) and the \textit{z $<$ 1 sample} (gray)}. 
The dotted lines indicate constant ratios of the two variables. Uncertainties caused by representative ranges of extinction ($A_V =\ $0 -- 2), outflow radial distance ($R_{out} =$ 7 -- 0.5 kpc), and electron density ($n_{e}=$ 1000 -- 100 cm$^{-3}$) are indicated by the orange, green, and blue bars, respectively. The values adopted in our calculations (0, 1 kpc, 200 cm$^{-3}$) are indicated by the squares. See \textit{Outflow Properties} in \textit{Methods} for more details.}
\label{fig:energetics}
\end{figure*}

These outflows are powerful based on order-of-magnitude estimates of their energetics (Extended Data Table \ref{tab:outflow}; see \textit{Outflow Properties} in \textit{Methods}).
They exhibit mass outflow rates of $\sim$ 140--46000 \msunyr. For J0807$+$1328, its mass outflow rate ($\sim$1000 \msunyr) is comparable to its SFR ($\sim$1082--1405 \msunyr\ \cite{Nguyen2020}).
While no reliable measurements of SFR are available for the remaining quasars in our sample, their mass outflow rates are comparable to or substantially higher than the typical SFR measured for z$\gtrsim$5 quasars ($\sim$100--3000 \msunyr) \cite{decarli_alma_2018,Nguyen2020,Wang2024}, suggesting a high efficiency in expelling gas out of these systems. 
Assuming a total gas mass of 10$^{11}$ \msun\cite{Fan2023}, these outflows may clear out the gas reservoir in $\sim$2 Myr--1 Gyr, consistent with the rapid quenching suggested by the abundance of post-starburst/quenched galaxies at z$\sim$3--5 \cite{Carnall2023,Glazebrook2024,deGraff2025,Valentino2023,Nanayakkara2024,Ji2026}. 
As shown in Fig. \ref{fig:energetics}, five objects from our sample show momentum outflow rates on the order of 10--200$\times$ the quasar radiation force (\Lbol/c), and only five objects have momentum outflow rates lower than \Lbol/c. Such significant momentum boost is broadly consistent with energy-conserving quasar-driven outflows which are capable of expelling gas out of the galaxies efficiently and might help with the formation of the M$_{\rm BH}$-$\sigma$ relation \cite{king15}. Five objects in our sample show kinetic energy outflow rates $\sim$10--260\% of \Lbol, clearly exceeding the thresholds ($\sim$0.1\%--5\%) required for the suppression of star formation activities within galaxies through quasar feedback based on various simulations \cite{Harrison2018}. Their kinetic energy outflow rates are comparable to or $\sim$1 order-of-magnitude larger than those of ERQs and $\sim$1--{4} order-of-magnitudes larger than those of {the \textit{z $\sim$ 1.5--3.5 sample} and \textit{z $<$ 1 sample}}, making them among the most energetic outflows known so far. While less powerful, the remaining outflows in our sample generally have kinetic energy outflow rates greater than both the median value of the {\textit{z $\sim$ 1.5--3.5 sample}} and 0.1\% of \Lbol, which can still provide negative feedback to some extent. {Collectively, the average (median) kinetic energy outflow rates of our sample is $\sim$315$\times$ ($\sim$26$\times$) higher than those of the \textit{z $\sim$ 1.5--3.5 sample} and $\sim$324$\times$ ($\sim$31$\times$) higher than those of the \textit{z $<$ 1 sample}}, suggesting dramatically enhanced feedback from luminous blue Type 1 quasars at z$\sim$5--6 than those at lower redshifts.

Overall, the high detection rate of extremely fast and powerful quasar outflows at z$\sim$5--6 draws a compelling picture where intense quasar feedback on galaxy scale is already at work just $\sim$1 Gyr after the Big Bang. They may be the long-sought strong evidence of quasar feedback responsible for the rapid quenching of the earliest massive post-starburst/quiescent galaxies. Furthermore, the suppression of stellar mass growth caused by such intense feedback may also help explain the overmassive BHs with respect to their host galaxies at z$>$5 when compared to local relations \citep{Fan2023,yue_eiger_2024}.

\captionsetup[table]{name=Extended Data Table}
\captionsetup[figure]{name=Extended Data Figure}

\newpage

\section*{Methods}\label{sec11}

\subsection*{Observation and Data Reduction}
\label{subsec:obs}

Our objects are observed with JWST NIRSpec Instrument in IFU mode \citep{Bok2022, Jak2022}  through the JWST cycle 2 survey program \#3428 (Q-IFU; PI W. Liu). The parent sample consists of 100 spectrally-confirmed bright quasars at z $\sim$ 4.7--4.9 and at z $\sim$ 5.6--6.0, with {absolute magnitude at rest-frame 1450 \AA\ M$_{1450}$ $\lesssim-$25.5}, representative of all luminous quasars at z$\sim$5--6 known so far. The two redshift ranges are chosen to place key emission features (\hb, \oiii, and optical iron emission features) and 5100\AA\ continuum within the spectral ranges of JWST observations. They are randomly chosen from the quasars published in \citep{sdssdrq12,Wang2016,Fan2023ext,Yang2016} to maximize the sky coverage. In the end, 28 objects are observed, which are randomly selected based on JWST scheduling. Among them, one object is heavily contaminated by artificial imprints from detector persistence and is unusable for our science analysis. The remaining 27 objects make up our final sample.

The sources at z $\sim$ 4.7--4.9 and z $\sim$ 5.6--6.0 are observed in configurations G235H/F170LP and G395H/F290LP, with corresponding wavelength coverage of 1.66--3.05 and 2.87–5.14~\mum, respectively. The gratings have a nominal resolving power $\lambda / \Delta\lambda \simeq 2700$, corresponding to a velocity resolution $\sim 110$ \kms, which allows us to spectrally resolve the emission lines well. The field-of-view of the IFU is 3\arcsec$\times$3\arcsec.
We use a 2-point small cycling dither pattern with 15 groups and 1 integration per position to improve the bad pixel rejection and a NRSIRS2RAPID readout pattern. The total on-source exposure is 466.8s per target. Although this setting is adequate for quasar spectroscopy, the poor spatial sampling and shallow exposure prevent us from determining the spatial extent of extended emission robustly.

The NIRSpec/IFU data are reduced following the general 3 stages of JWST\ Science Calibration Pipeline (version ``1.14.0'' and context file ``jwst\_1293.pmap''), combined with customized software and scripts to replace or improve certain steps in the public pipeline and produce the final data cube properly. Specifically, after stage 1, we further subtract the correlated detector noise (1/$f$ noise) in the count rate images using NSCLean \citep{NSclean}, where the correlated vertical noise in each column (i.e. along the spatial axis) is modeled with a 2nd-order polynomial function, after all bright pixels associated with the observed target have been removed through sigma-clipping. After stage 2, we apply further sigma-clipping in each calibrated 2D spectral images to get rid of outliers \cite{Vayner2023b}. The final IFU data cube is reconstructed with ``drizzle'' method adopting a spatial pixel (spaxel) size of 0\arcsec.1.
A master background is built by integrating over spaxels covering blank sky and subtracted from the data cube.
For each object, a spatially-integrated spectrum is extracted from each IFU data cube with a r$=$0\arcsec.3 circular aperture. The only exceptions are J0807$+$1328, J0859$+$2520, J1458$+$3327, where r$=$0\arcsec.5, 0\arcsec.1, and 0\arcsec.2 apertures are used instead to maximize the S/N of their spectra. {The artificial spectral oscillations known to be significant in individual IFU spaxels  \cite{Law2023} are negligible in these spectra ($<$0.5\% in general), especially when compared to the strong quasar emission}.
Finally, we apply an aperture correction to each spectrum based on the curve of growth analysis results from the NIRSpec/IFU data of the standard star TYC 4433-1800-1 (program ID: 1128) over the same wavelength range, which are 10.2--140.0\% (10.2--11.0\% excluding J0859$+$2520) for G235H observations and 11.3--20.4\% for G395H observations, respectively.

\subsection*{Comparison Samples}
\label{subsec:control}

\begin{figure}[!htb]
\centering
\includegraphics[width=0.5\textwidth]{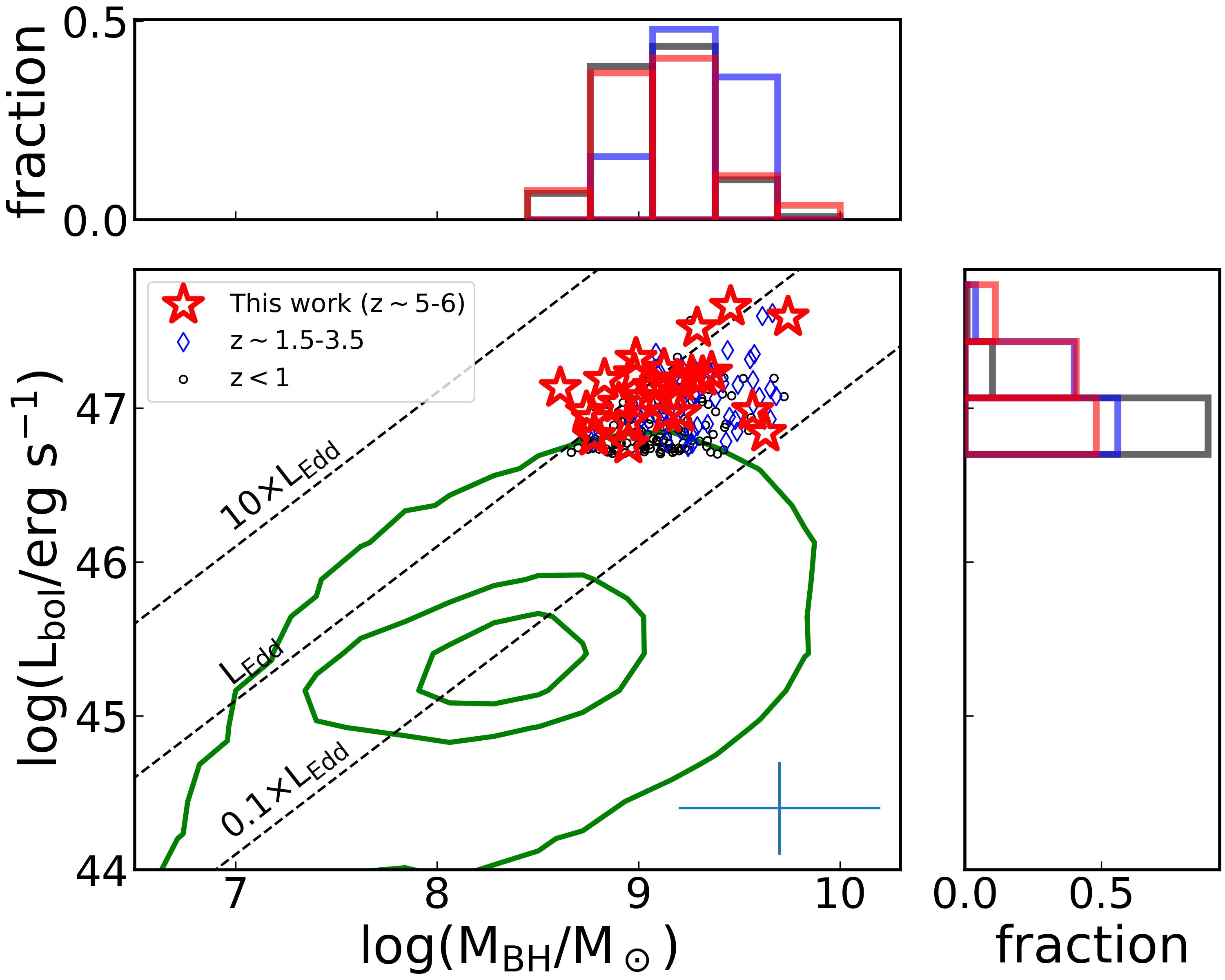}
    \caption{\textbf{Quasar properties.} \Lbol\ versus \mbh\ for our sample (red), the \textit{z$\sim$1-3 sample} (blue) and the \textit{z $<$ 1 sample} (black). The green contours represent the distributions of 99.998\%, 90\% and 50\% of all z $<$ 1 SDSS quasars \cite{Wu2022ext}. The dashed lines show the locations of constant accretion rates at 0.1, 1 and 10 times the Eddington luminosity. The typical statistical uncertainties for the \Lbol\ and \hb-based \mbh\ are indicated by the blue cross. The top and right panels show the fraction histograms for the corresponding samples with the same colors.}
    \label{fig:Lbol_MBH}
\end{figure}

{Our sample exhibits rest-frame optical spectral properties closely resembling those of lower-redshift luminous Type 1 quasars, except for the \oiiitext\ profiles tracing outflows.}
{Our sample and the other two comparison samples are all selected using a very similar color selection technique based on the characteristic SED of Type 1 quasars with prominent blue, power-law continuum \cite{Richards2002}. Our objects are selected from either the SDSS DR16Q catalog \cite{DR16Q} or Pan-STARRS1 surveys \cite{panstarr} as compiled by \cite{Fan2023ext}.
The two comparison samples are both selected from the SDSS DR16Q catalog \cite{DR16Q,Wu2022ext}. 
The \textit{z $\sim$ 1.5--3.5 sample} is made up of all quasars with rest-frame optical spectra available from \cite{Shen2016ext} and \Lbol\ \mbh, and \edd\ falling within the same ranges as those of our z$\sim$5--6 sample. For quasars in \cite{Shen2016ext}, the only selection criteria are that their SDSS optical spectra have S/N per pixel $\gtrsim$ 10 and that they fall in the appropriate redshift windows to allow for observations of the \hb--\oiiitext\ region with ground-based spectroscopy. The \textit{z $<$ 1 sample} is made up of all quasars at z $<$ 1 with \Lbol, \mbh, and \edd\ falling within the same ranges. The distribution of \mbh\ and \Lbol\ of the three samples are shown in Extended Data Fig. \ref{fig:Lbol_MBH}, demonstrating that they all represent the same population of luminous quasars. 
}

{To further compare our sample and the two comparison samples, we measure their median and standard deviations of \Lbol, \mbh\ and \edd. For the median values, we use the Monte Carlo method to estimate their 90\% confidence levels, where we assume that the uncertainties of individual measurements follow a normal distribution with a standard deviation of 0.3 dex for \Lbol\ and 0.5 dex for \mbh\ and \edd. Here the standard deviations are set as the typical statistical uncertainties for corresponding quasar properties. \cite[e.g.][]{Shen2013}. The obtained results are shown in Extended Data Table \ref{tab:comparison}, which suggest that the differences of the \Lbol, \mbh\ and \edd\ among the three samples are insignificant.} 

{The higher detection rate of extreme outflows in our sample is not caused by such small differences in the quasar properties among the three samples. Considering all objects with \oiiitext\ detections in the three samples combined, the \oiiitext\ velocity \vjiuba\ does not correlate with the \Lbol, \mbh\ and \edd: Based on the Kendall's tau tests \cite{Kendallext}, the corresponding $p$-values are 0.96, 0.28 and 0.24, respectively, with the null hypothesis of no correlation. Moreover, even if we choose objects within narrower \Lbol, \mbh\ and \edd\ ranges so that the quasars properties of the three samples overlap even better with each other, the statistical conclusions are still similar despite of the smaller sample sizes. For example, if we only consider objects within $\pm{0.3}$ dex of the median log$(L_{\rm bol}$/erg s$^{-1}$) (46.9), log(M$_{\rm BH}$/\msun) (9.1) and \edd\ (0.5) of the two lower-redshift samples combined, there will be 2/9 (22.2$^{+28.5}_{-13.5}$\%) extreme outflows in our sample, 
whereas 1/104 or 1.0$^{+3.5}_{-0.6}$\% in the two lower-redshift samples combined.}

{
Another factor that may affect the detection rate of extreme outflows is the orientation of the accretion disk with respect to the line-of-sight (LoS). Past studies \cite{Risaliti2011,Bisogni2017,Vietri2018} suggested that the \oiiitext\ equivalent width (EW) are systematically smaller in the face-on viewing angles where large-scale outflows are easier to be observed. This implies relatively small \oiiitext\ EW for our objects with extreme outflows in our sample.
Nevertheless, the observed \oiiitext\ EW of these objects are quite large instead ($\sim$ 26--112 \AA). In addition, the median \oiiitext\ EW of our sample ($\sim$10 \AA) is comparable to those of the \textit{z $\sim$ 1.5--3.5 sample} ($\sim$13 \AA) and \textit{z $<$ 1 sample} ($\sim$7 \AA).
Therefore, the higher fraction of extreme outflows in our sample is not caused by the orientation effect that our objects are more face-on than the two lower-redshift samples.}
Finally, we have confirmed that the S/N of spectra in our sample is comparable to those of \textit{z $\sim$ 1.5--3.5 sample} and \textit{z $<$ 1 sample} (with a mean of {$\sim$88, 88 and 100} in the rest-frame 5090--5110 \AA, respectively), which thus does not affect our results, either. {The spectra of all these quasar samples are fit with the same approach as described below. Overall, we expect no selection bias that would significantly change our results}.

\subsection*{Spectral Fitting}
\label{subsec:specfit}
For each object, we adopt the public software, \textit{PyQSOFit} \citep{pyqsofit} to model the spectrum and derive parameter uncertainties using MCMC. Specifically, the quasar pseudo-continuum is fit with a power law and empirical iron templates \cite{BorosonGreen1992,Vestergaard2001}, using the continuum windows free of other strong emission lines. The emission line-only spectrum is then obtained by subtracting the best-fit pseudo-continuum from the original spectrum. 
{To evaluate the uncertainties caused by the modelling of iron emission in our fits, we have also adopted the other two major iron templates \cite{VC04,Kovacevic2010} to fit the full wavelength ranges available for our data, in addition to our default iron template and fitting wavelength ranges. 
Overall, we find that our results are not significantly affected by the choice of iron templates and fitting wavelength ranges. The obtained \vjiuba\ of \oiiitext\ are $\sim$0.7--2.0$\times$ (median 1$\times$) those of our default fits. This suggests that, in general, the outflow velocities of our objects are not significantly overestimated in our best-fits and our main conclusions remain unaffected.}

For emission lines, the \hb\ and \hg\ are fit with up to 3 broad Gaussian components and 1 narrow Gaussian component. The kinematics (i.e., velocity and velocity dispersion) of each corresponding broad Gaussian component in the two lines are tied together. The \oiiitext\ doublet is fit with up to 2 Gaussian components where the kinematics of the narrow Gaussian component is tied to those of the narrow Gaussian components of \hb\ and \hg. The only exception is J0859$+$2520 where the narrow \oiii\ component is highly blueshifted with respect to \hb\ and the kinematics of the two are untied.
The \oiiitext\ doublet flux ratio of each Gaussian component is fixed at 1:2.98 \citep{Osterbrock2006} in the fit.  
The number of Gaussian components adopted in the final best-fit model is determined based on Bayesian information criterion (BIC) by choosing the model with the lowest BIC. As an example, the best-fit spectra and \oiii\ profiles for the 6 extreme outflows are shown in Fig. \ref{fig:profile} and Extended \ref{fig:samples}. 

We detect \oiiitext\ emission lines (with S/N $>$ 3) in 16 objects.
For each of them, we determine the systemic redshift adopting the narrow component of \oiiitext\ emission line, except for J0859$+$2520, J0941$+$5947, J1116$+$5833, J1342$+$5838, J1436$+$5007, J1458$+$3327. For J0859$+$2520, even the narrow \oiiitext\ emission line component is highly blueshifted with respect to \hb. For the remaining 5 objects, their best-fit \oiiitext\ models only contain broad components with FWHM $\gtrsim$ 1300 \kms\ and trace outflowing/turbulent gas. We thus use the peak of the overall \hb\ profiles to determine their redshifts. For the remaining objects with no \oiiitext\ detections, their redshifts are also determined using the peak of the overall \hb\ profiles. The final results are listed in Extended Data Table \ref{tab:outflow}. In addition, quasar J1620$+$5202 shows the broadest and most blueshifted \oiiitext\ emission lines. It is likely that all emission lines are blueshifted with respect to the true systemic velocity and the redshift based on the narrowest \oiiitext\ is an underestimate. 
{In comparison, we detect \oiiitext\ emission lines in 31/50 ($\sim$62\%) objects of the \textit{z $\sim$ 1.5--3.5 sample} and 92/119 ($\sim$77\%) objects of the z $<$ 1 sample. They are similar to or moderately larger than the \oiiitext\ detection rate of our sample ($\sim$59\%). The detection rates of extreme outflows in objects with \oiiitext\ detections are still significantly higher in our sample (6/16 or $\sim$38\%) than in the two lower-redshift samples (0/31 or 0\% and 3/92 or $\sim$3\%).}

\subsection*{Nuclear Properties}
\label{subsec:nuclear}

The \Lbol\ of our objects are derived adopting 5100\AA\ continuum luminosities (\Lwu) with 
 a bolometric correction factor of 9.26 \cite{Richards2006}. The results are listed in Extended Data Table \ref{tab:outflow}.
The \mbh\ of our objects are obtained from the scaling relation
in \cite{Vestergaard2006}:

\begin{equation}
\begin{split}
{\rm log}(M_{\rm{BH}}) = {\rm log}\left\{\left[\frac{{\rm FWHM}(\rm{H\beta})}{1000\ {\rm km\ s}^{-1}}\right]^2 \left[\frac{\lambda L_\lambda(5100 )}{10^{44}\ {\rm erg\ s}^{-1}}\right]^{0.50}
\right\} + (6.91 \pm{0.02})
\end{split}
\label{eq:BH}
\end{equation}

Here FWHM(\hb) are calculated for the overall line profile of broad \hb\ emission line, without subtracting the potential contribution from outflowing gas. {Note that when modeling and removing an outflow component sharing the same kinematics as that of \oiiitext\ from \hb, the changes for FWHM(\hb) are in general $<$15\% and the corresponding changes in $M_{\rm BH}$ are $<$23\%. This is smaller than the typical statistical uncertainty of \hb-based \mbh\ (0.5 dex) and thus does not affect our \mbh-related results significantly.} The Eddington ratios (\edd) are then derived adopting $L_{\rm Edd} = 1.26\times10^{38} (M_{\rm BH}/M_\odot)$ erg s$^{-1}$. 
The individual values of \Lbol, \mbh\ and \edd\ are presented in \cite{Liu2025bb}.

\subsection*{Velocity Thresholds for Outflows and Extreme Outflows}

{In this study, while we broadly classify all objects with 50th percentile velocity \vwu\ $<$ 0 \kms\ as outflows, only those with \vwu\ $<-$200 \kms\ (9 objects in our sample) have robust outflows, since such velocities are larger than the typical line width of normal high-redshift galaxies. Objects with $-$200 \kms\ $<$ \vwu\ $<$ 0 \kms\ (7 objects in our sample) are more uncertain: they have modest LoS velocities which could be attributed to non-outflowing gas, but they could still have large outflow velocities outside of the LoS. Nevertheless, the exact velocity threshold for outflows does not change our main conclusions which depend on the fraction of extremely fast outflows with much larger velocities.}

To estimate the escape velocities of our objects, we adopt 10$^{12.4}$--10$^{13}$ \msun\ dark matter halos with Navarro–Frenk–White (NFW) \cite{NFW} profiles and concentration parameters from \cite{DuttonMaccio2014}, using the software \textit{galpy}\cite{galpy}. We obtain an escape velocity of $\sim$570 -- 900 \kms. The lower dark matter halo mass corresponds to the recent estimates of typical value for z$\sim$6 quasars \cite{Costa2024,Wang2023,Eilers2024} and the higher value corresponds to that of a typical early type galaxy\cite{Wechsler2018}. {To ensure that our definition of extreme outflows can be applied to even the \textit{z $<$ 1 sample} which may have elliptical host galaxies in many cases \cite{KormendyHo2013}, we choose to use the higher value (900 \kms) as the escape velocity in our calculation.}
{For a Gaussian profile, $|$\vjiuba$|$ $=$ $|$\vwu$|$ $+$ 2$\sigma$. If we define an extreme outflow as the one with \vwu\ $>$ \vesc\ and $\sigma$ $>$ \vesc, then it leads to $|$\vjiuba$|$ $>$ 3\vesc. We adopt this as our empirical threshold for extreme outflows in this study, even though the emission line profiles are generally non-gaussian in our objects. 
Nevertheless, the exact threshold adopted to define extreme outflows does not affect our results in general: As shown in Fig. \ref{fig:hist}, at a given velocity v0, the fraction of objects with \vjiuba\ $<$ v0 in our sample is always larger than the two lower-redshift samples, all the way up to a v0 of $\sim-$1500 \kms.}

\subsection*{Outflow Properties}
\label{subsec:dynamics}

\begin{figure*}[!h]
\centering

\begin{minipage}[t]{0.32\textwidth}
\includegraphics[width=\textwidth]{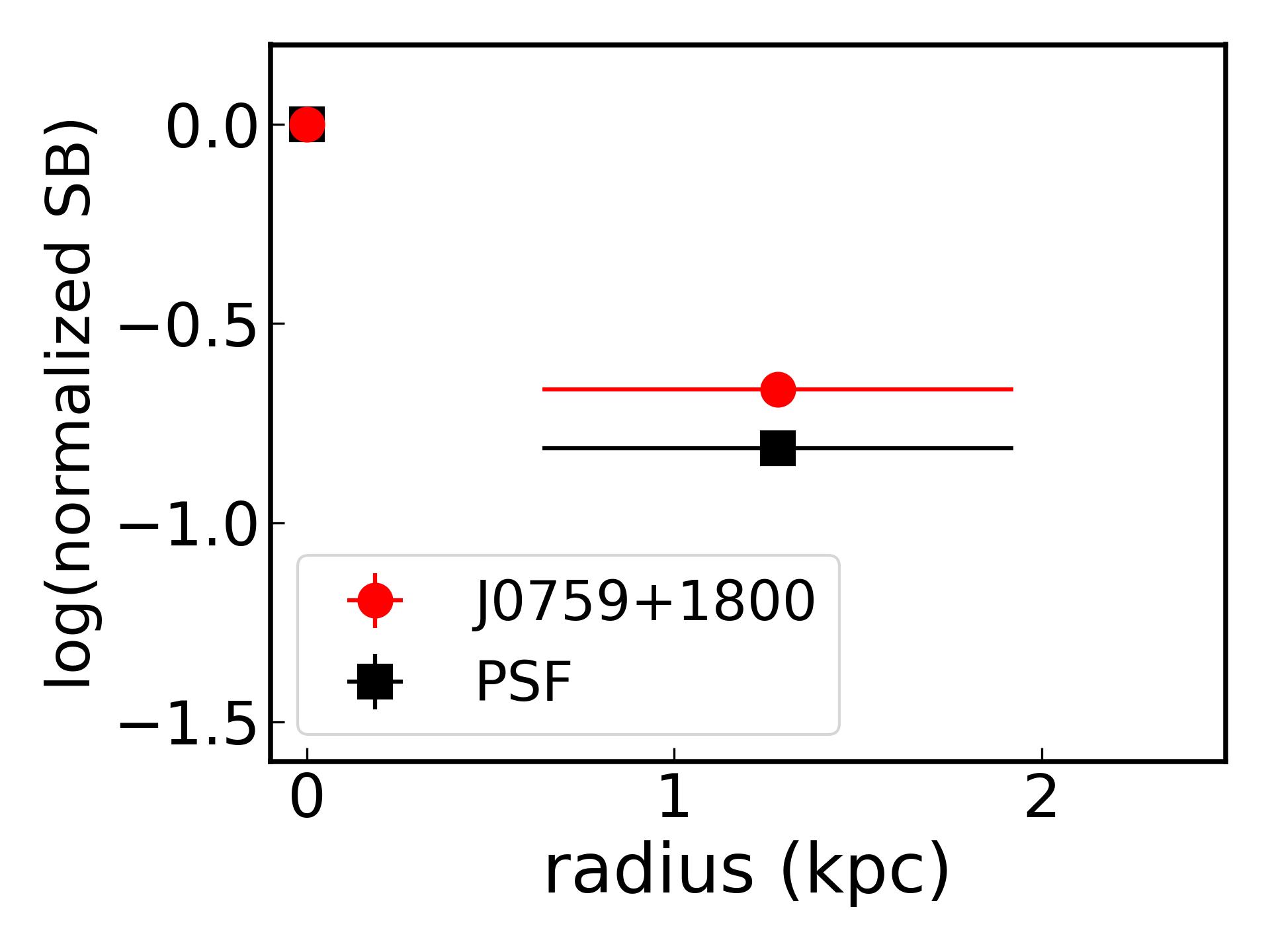}
\end{minipage}
\begin{minipage}[t]{0.32\textwidth}
\includegraphics[width=\textwidth]{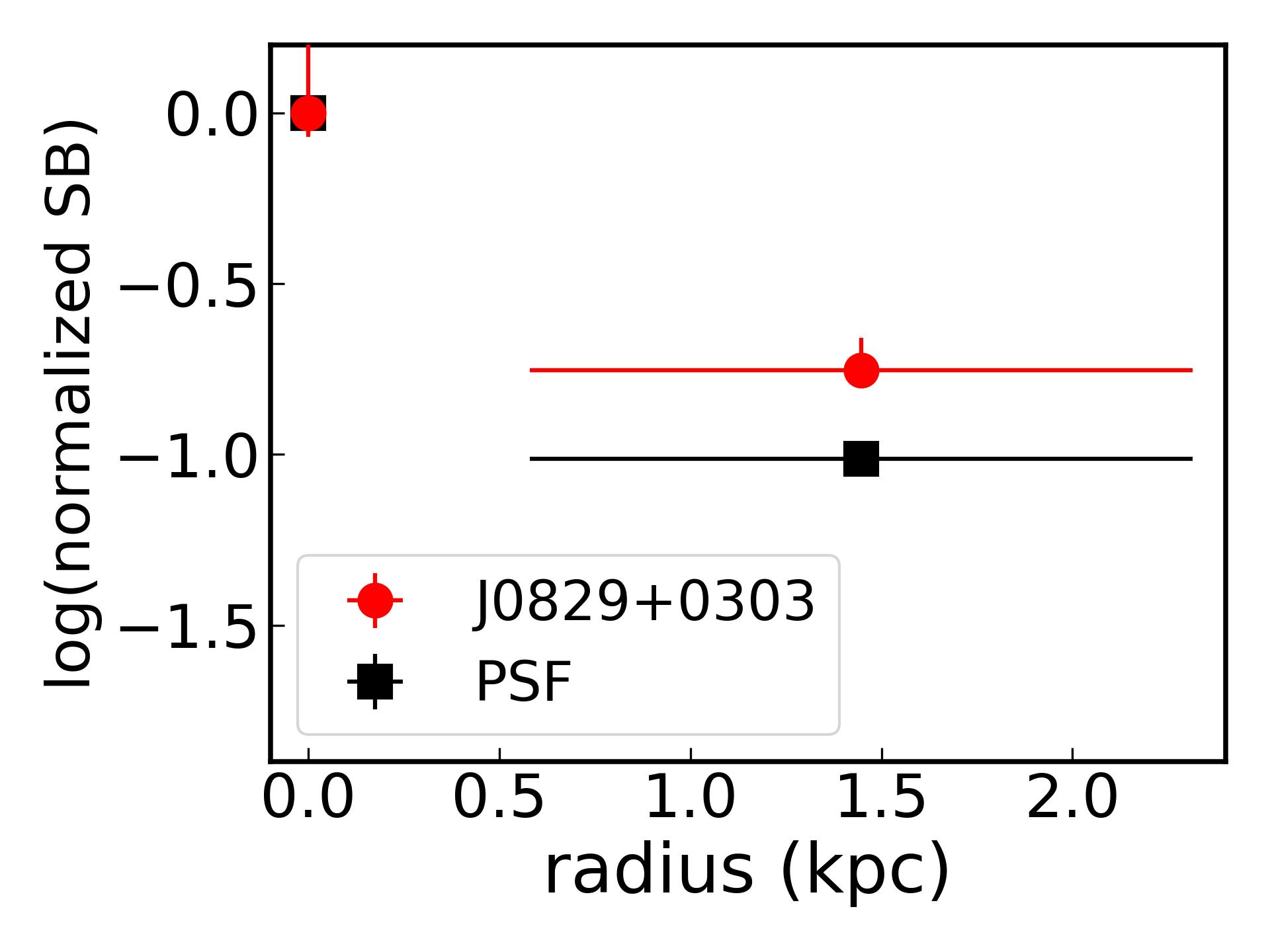}
\end{minipage}
\begin{minipage}[t]{0.32\textwidth}
\includegraphics[width=\textwidth]{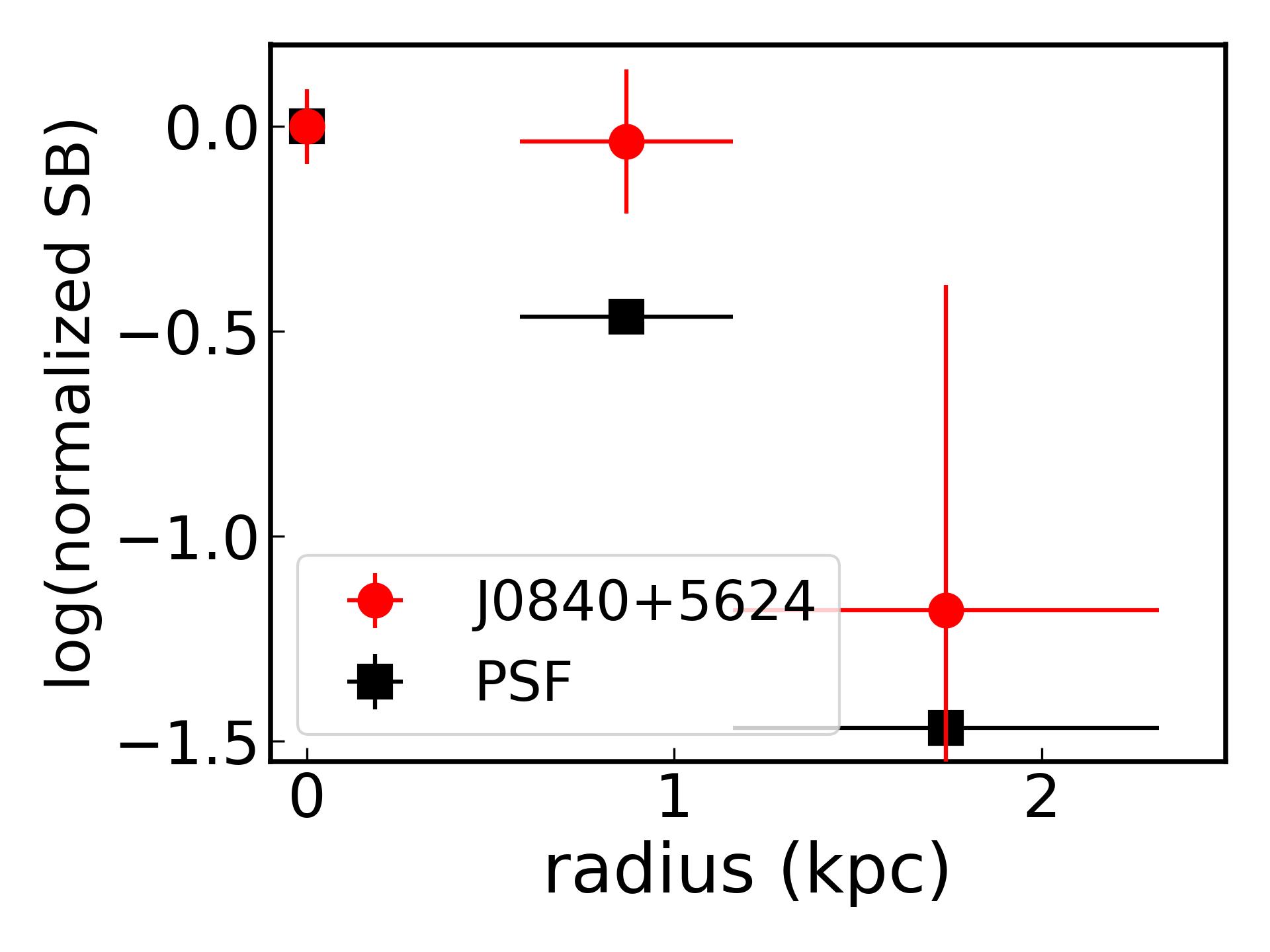}
\end{minipage}

\begin{minipage}[t]{0.32\textwidth}
\includegraphics[width=\textwidth]{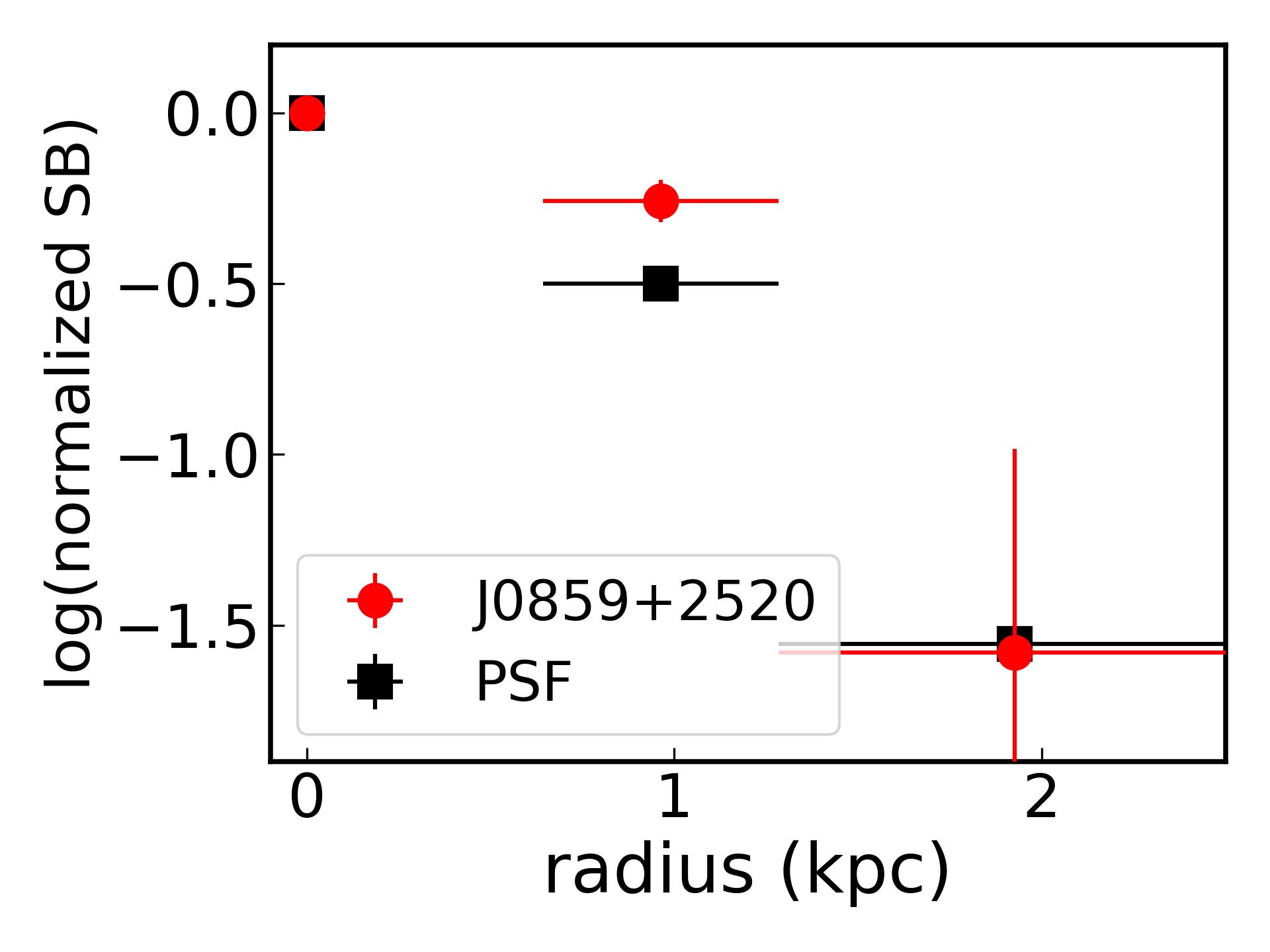}
\end{minipage}
\begin{minipage}[t]{0.32\textwidth}
\includegraphics[width=\textwidth]{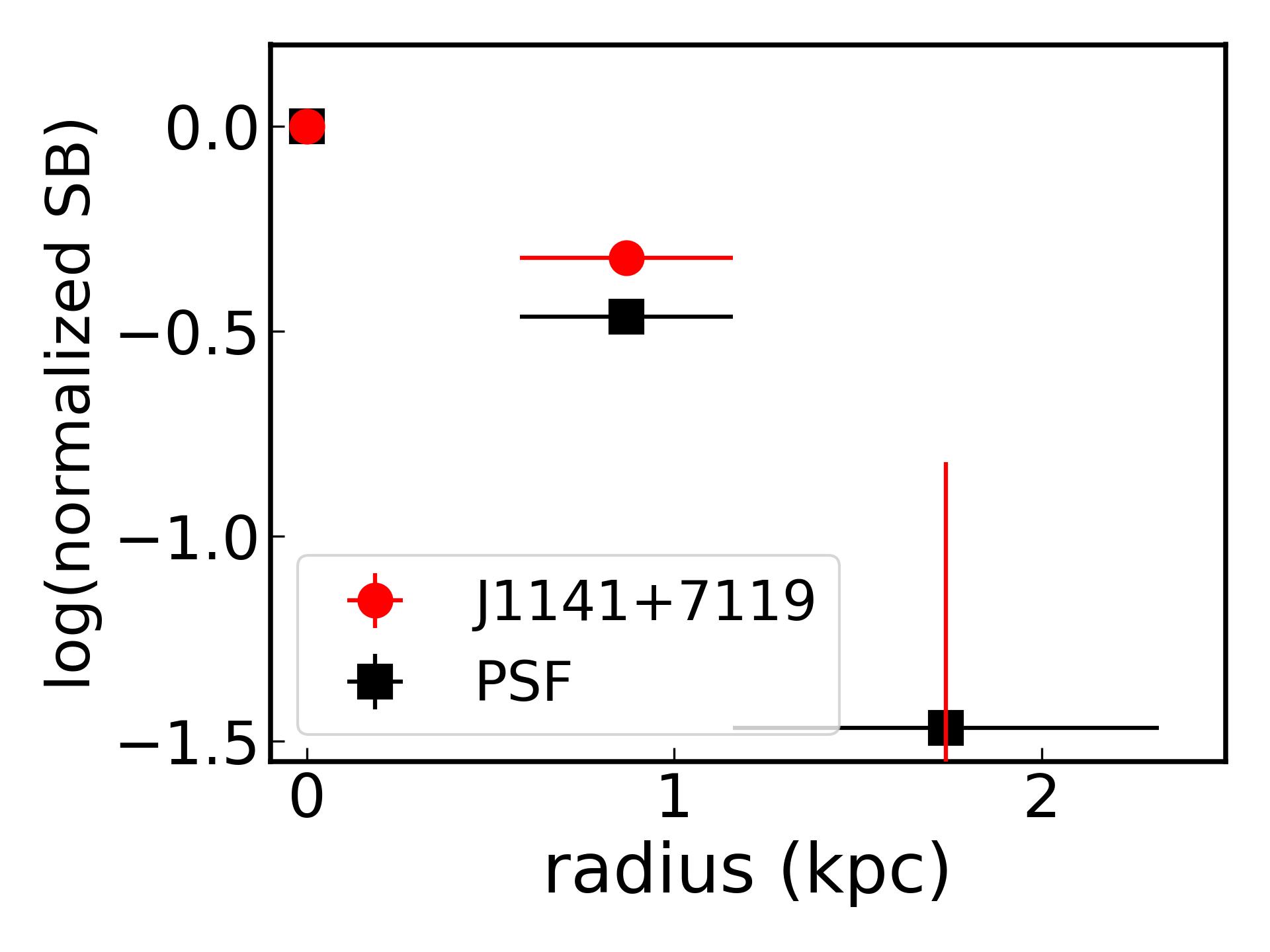}
\end{minipage}
\begin{minipage}[t]{0.32\textwidth}
\includegraphics[width=\textwidth]{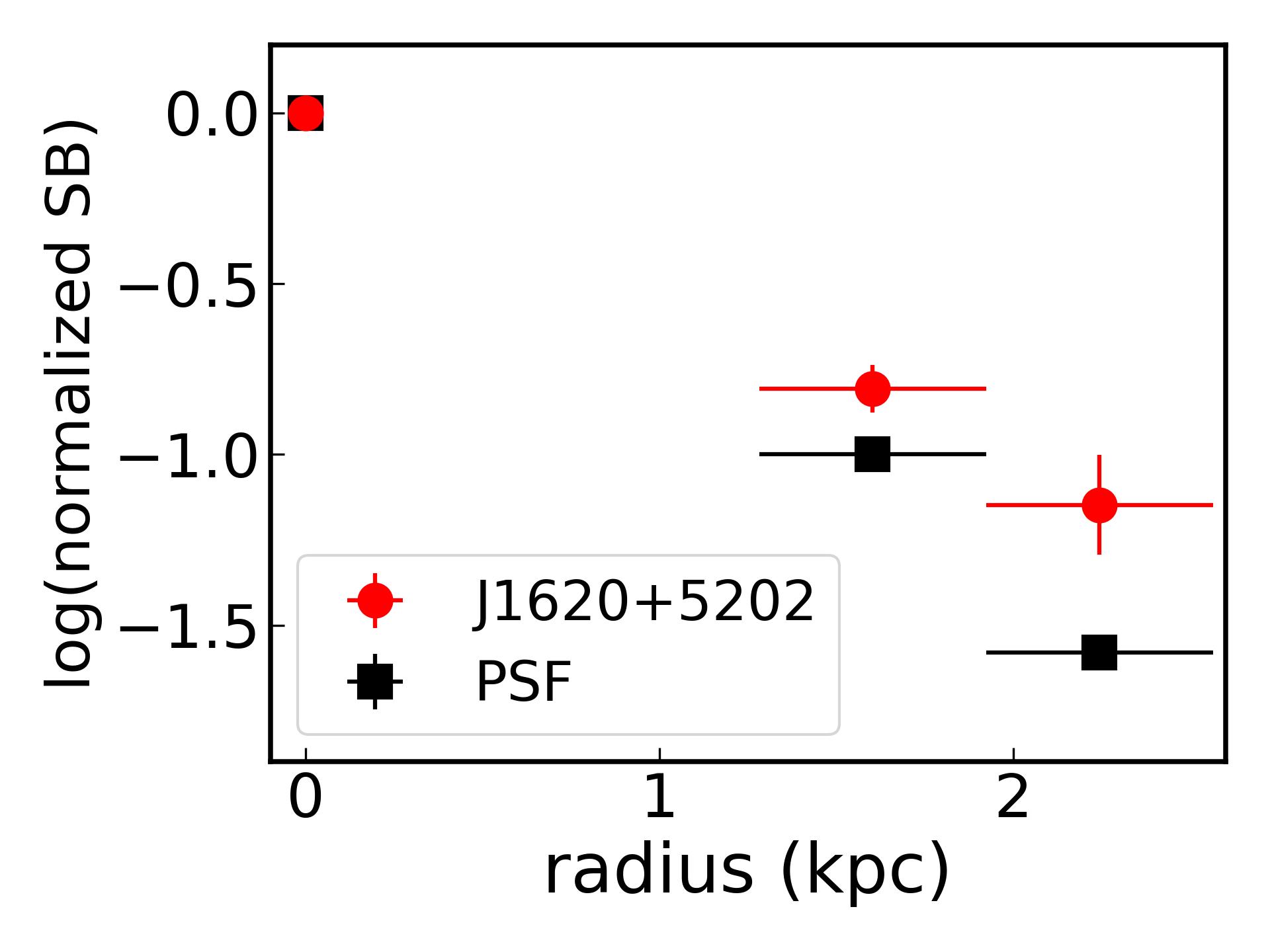}
\end{minipage}

\caption{\textbf{Radial profiles of outflows}. Azimuthally-averaged and normalized radial surface brightness (SB) profiles for the 6 extreme outflows in our sample (red) versus the corresponding PSF. The horizontal error bars indicate the sizes of the radial bins. See \textit{Outflow Properties} for more details.}
    \label{fig:spatial2}
\end{figure*}

The extreme outflows in our sample are on galaxy scale (kpc scale) in general. In Extended Data Fig. \ref{fig:spatial2}, we show the azimuthally-averaged and normalized radial surface brightness profiles of the outflowing gas, with flux integrated over the [$-$4000, $-$1000] \kms\ range of the \oiii\ emission line ([$-$8000, $-$1000] \kms\ for J1620$+$5202, the fastest outflow in our sample). The radial bin sizes, with values of 0.1\arcsec, 0.2\arcsec\ or 0.3\arcsec, are optimized for individual objects to enhance S/N, which leads to a limited number of bins for each object. For each radial bin, the \oiiitext\ flux of the outflowing gas is measured as the difference between those from two circular apertures with radii matching the inner and outer boundaries of the radial bin. The \oiiitext\ flux within each aperture is obtained from the best-fit of corresponding spectrum modeled with the same approach described in \textit{Spectral Fitting}. For each object, the PSF profile is built in the same way over the corresponding wavelength range from the IFU data cubes of the standard star (TYC 4433-1800-1; program ID: 1128). 
All 6 extreme outflows in our sample are more extended than PSF on a scale of $\sim$ 1 kpc, confirming that they are on galaxy scale. Since outflows in such quasars are usually not spherically symmetric \cite{Liu2024bext}, the azimuthally-averaged radial profiles adopted here will actually weaken the deviation of the outflow emission from the PSF. Therefore, our results are conservative, reinforcing that these outflows are extended on galaxy scale.

We adopt a widely-used non-parametric approach to describe the \oiii\ line profile and measure the {outflow velocities} \cite{zaka14, Perrotta2019ext}. Specifically, {\vjiuba, \vjiu\ and \vwu\ are the velocities smaller than those representing 98\%, 90\% and 50\% of the line flux}, respectively.
\wjiu\ is the line width enclosing 90\% of the total line flux between \vjiuwu\ and \vlingwu, the velocities where 95\% and 5\% of the line flux have larger velocities. These results are listed in the Extended Data Table \ref{tab:outflow}. These non-parametric kinematics measurements only depend on the overall emission line profiles and are thus insensitive to the fitting details (e.g., numbers of Gaussian components adopted in the best-fits). This allows for impartial comparisons with outflows discovered in other quasar samples.

Next, {for all quasar samples}, we calculate the outflowing gas mass ($M_{\rm out}$) and time-averaged mass ($\dot{M}_{\rm out}$), momentum ($\dot{p}_{\rm out}$) and kinetic energy outflow rates ($\dot{E}_{\rm out}$) following \cite{Liu2024bext,Liu2025}:

\begin{eqnarray}
M_{\rm out} & = & 5.3 \times 10^8~\frac{C_e L_{44}([\rm{O~{III}}])}{n_{e,2} 10^{[\rm O/H]}} M_\odot\\
\dot{M}_{\rm out} & = & M_{\rm out}~(v_{\rm out}/R_{\rm out}) \\
\label{eq:pdot}
\dot{p}_{\rm out} & = & \dot{M}_{\rm out}~v_{out} \\
\dot{E}_{\rm out} & = & \frac{1}{2}~\dot{M}_{\rm out}(v_{\rm out})^2 
\end{eqnarray}

Here {$C_e \equiv \langle n_e \rangle^2 / \langle n_e^2 \rangle $} is the electron density clumping factor and should be {of order unity} based on a cloud-by-cloud basis (i.e. each gas cloud of the outflowing gas has uniform density). $L_{44}$ is the \oiii\ luminosity in units of 10$^{44}$ erg s$^{-1}$. To facilitate fair comparisons among all samples, 
we derive outflow energetics consistently with the same approach as adopted in literature \cite{Perrotta2019ext}, including: 

\textbullet\ We use the total \oiiitext\ luminosity in our calculations when two Gaussian components are required for the best-fits. If we instead use the luminosity of the broader \oiiitext\ component as conservative lower limits, the total mass will be $\sim$20\%--70\% (median $\sim$30\%) lower. Note that they are lower limits since the narrower components are usually quite broad ($>$1000 \kms) in our objects and a significant part of them are still likely outflowing gas. We apply no extinction correction to the \oiiitext\ luminosity as there is no clear evidence of dust reddening in the quasar continuum. Nevertheless, non-negligible dust extinction cannot be ruled out for the outflowing gas based on our current data. A$_V$ up to $\sim$2 have been reported in extended gaseous nebulae in quasars at z$\sim$6 \cite{Decarli2024ext}.
{Moreover, the outflows in ERQs may have larger dust-extinction corrections, given the dusty environments within them.}

\begin{figure}[!htb]
    \centering
    \includegraphics[width=0.5\linewidth]{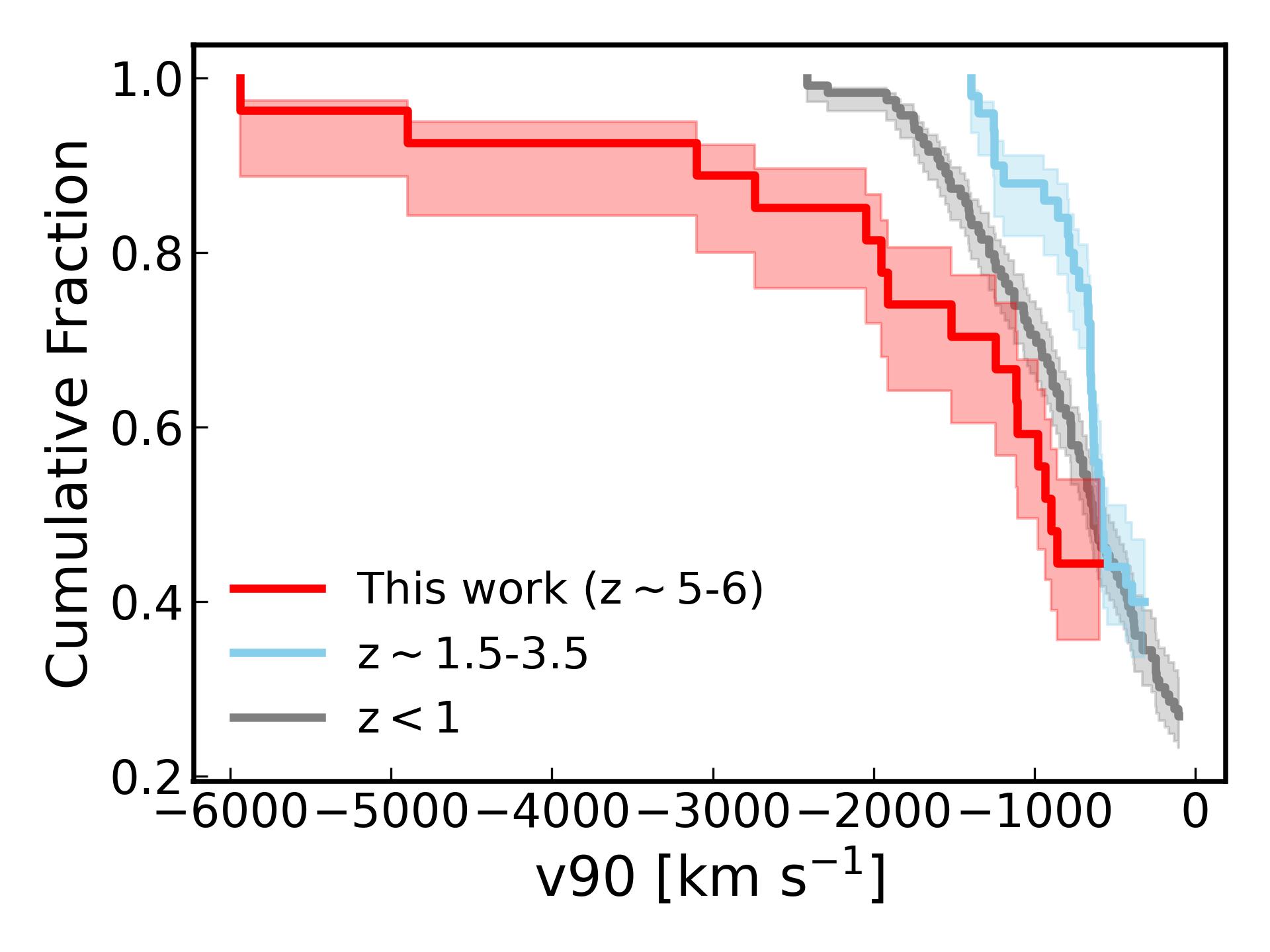}
    \caption{\textbf{Cumulative distribution function of \vjiu.} Same as the bottom panels of Fig. \ref{fig:hist} but for the \vjiu\ of \oiiitext.}
    \label{fig:v90}
\end{figure}

{\textbullet\ We define the outflow velocity $v_{out}$ as \vjiuba\ following previous studies (e.g., \cite{Rupke13a,Perrotta2019ext}), which represents the maximum speed of the outflow, or the lower limit of it if the outflow is mostly outside of the LoS.} In addition to the line centroid shift, it also encodes the emission line width and therefore accounts for outflow velocity outside of the LoS and turbulent motions, whereas adopting \vwu\ underestimates the true outflow velocity in 3D space due to projection effect. {Previous studies also adopt slightly different outflow velocities and may reduce the absolute outflow energetics. For example, if we adopt \vjiu\ as $v_{out}$ instead, the $\dot{M}_{\rm out}$, $\dot{p}_{\rm out}$ and $\dot{E}_{\rm out}$ will decrease by $\sim$23\%--45\%, $\sim$40\%--70\%, and $\sim$54\%--84\%, respectively. Nevertheless, our primary results that extreme outflows are more frequent and on average more powerful in our sample is not sensitive to the definition of outflow velocity. For example, the \vjiu\ cumulative distribution function of our sample is still different from those of the two lower-redshift samples as shown in Extended Data Fig. \ref{fig:v90}. The Mann-Whitney \textit{U} tests give $p$-values $\sim$ 5.1$\times$10$^{-5}$ and $\sim$6.7$\times$10$^{-4}$, respectively.}

\textbullet\ The outflow radial distance R$_{out}$ is assumed to be 1 kpc following \cite{Perrotta2019ext}, which is supported by the evidence that the outflows are spatially extended on a scale of $\sim$1 kpc in our objects (see Extended Data Fig. \ref{fig:spatial2}).
This is comparable to the typical values ($\lesssim$1--4 kpc) revealed by spatially resolved studies of quasar-driven fast \oiiitext\ outflows at similar cosmic epoch \cite{Marshall2023ext,Marshall2025,Liu2024bext,Vayner2024}. Moreover, typical ranges of outflow radial distance adopted in other lower-redshift outflow studies are 0.5-7 kpc \cite{Bischetti2017,Perrotta2019ext}, again consistent with the value we adopt. 

\textbullet\ The electron density $n_e$ is assumed to be 200 cm$^{-3}$ (or electron density in units of 100 cm$^{-3}$, $n_{e,2}$$=$2) and the metallicity  is assumed to be solar ($[\rm O/H]$$=$0). These properties cannot be robustly measured from our data. The $n_e$ adopted is the same as those used in other studies of outflows in luminous quasars \cite{Bischetti2017,Mingozzi2019,Perrotta2019ext} and close to that measured from the outflow in a z$\sim$6 quasar based on \sii\ doublet ratio (300--500 cm$^{-3}$)\cite{Decarli2024ext}. Furthermore, the typical range of electron density for lower-redshift quasar outflows reported in the literature is 100 -- 1000 cm$^{-3}$ \cite{Liu2013b, Harrison2014,Perrotta2019ext}. 
In the same quasar with \siitext-based $n_e$ as mentioned above, the gas-phase metallicity of the host galaxy is close to solar ($\sim$0.6--0.8 Z$_\odot$), despite the systematic uncertainties caused by the uncalibrated empirical relations. On the other hand, the mass-metallicity relation at z$>$5 suggests solar metallicity at stellar mass over 10$^{10}$ \msun \cite{Nakajima2023}, comparable to or less massive than the expected mass of quasar host galaxies in our sample (based on the average BH to stellar mass ratio in z$\sim$6 quasars \cite{yue_eiger_2024ext}). These provide circumstantial evidence that the metallicity of outflowing gas in our sample should not be significantly lower than the solar value.

The impact of typical uncertainties of the adopted parameters to the resulting outflowing energetics,  due to the extinction correction (A$_V$$=$0--2), outflow radial distance (7--0.5 kpc) and electron density (1000--100 cm$^{-3}$), are shown in Fig. \ref{fig:energetics}. Note that these uncertainties may change the outflow energetics in opposite directions: a non-zero dust extinction correction increases the outflow energetics whereas larger electron density and radial distance decrease it. Overall, given the assumptions listed above, the outflow energetics for our objects should be treated as orders-of-magnitude estimates. Nevertheless, the differences between our sample and the other quasar samples are significant, since their outflow energetics are derived under the same assumptions. Deep, fully spatially-resolved JWST NIRSpec/IFU observations are required to derive much more precise outflow energetics for our objects. Together with future observations on the star formation, cool gas and stellar properties of their host galaxies, they will provide more insight on the impact of these outflows.

\backmatter

\newpage

\section*{Extended Data}

\begin{figure}[!h]
\setcounter{figure}{0}
\begin{minipage}[t]{0.5\textwidth}
\centering
\includegraphics[width=\textwidth]{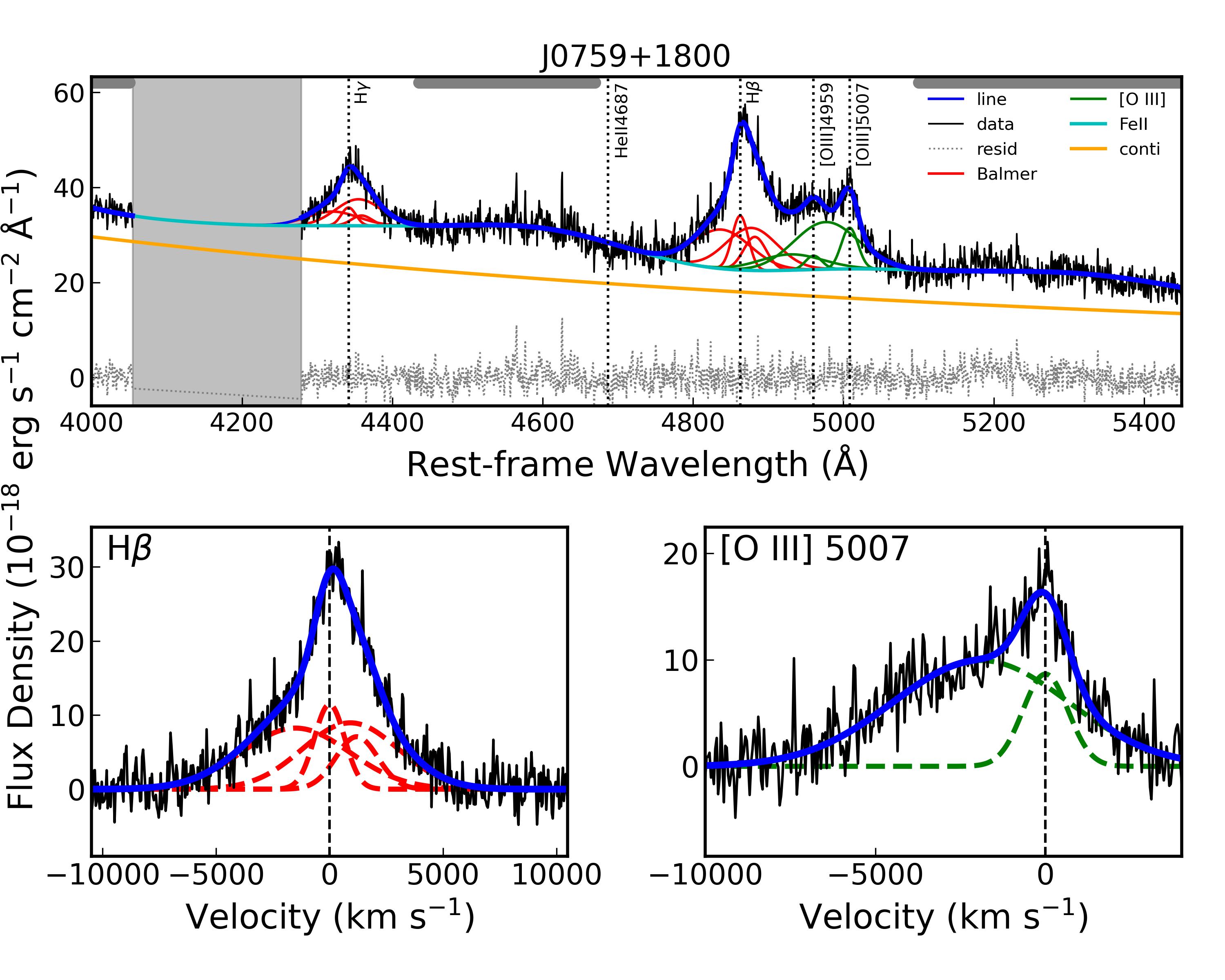}
\end{minipage}
 \begin{minipage}[t]{0.5\textwidth}
 \centering
\includegraphics[width=\textwidth]{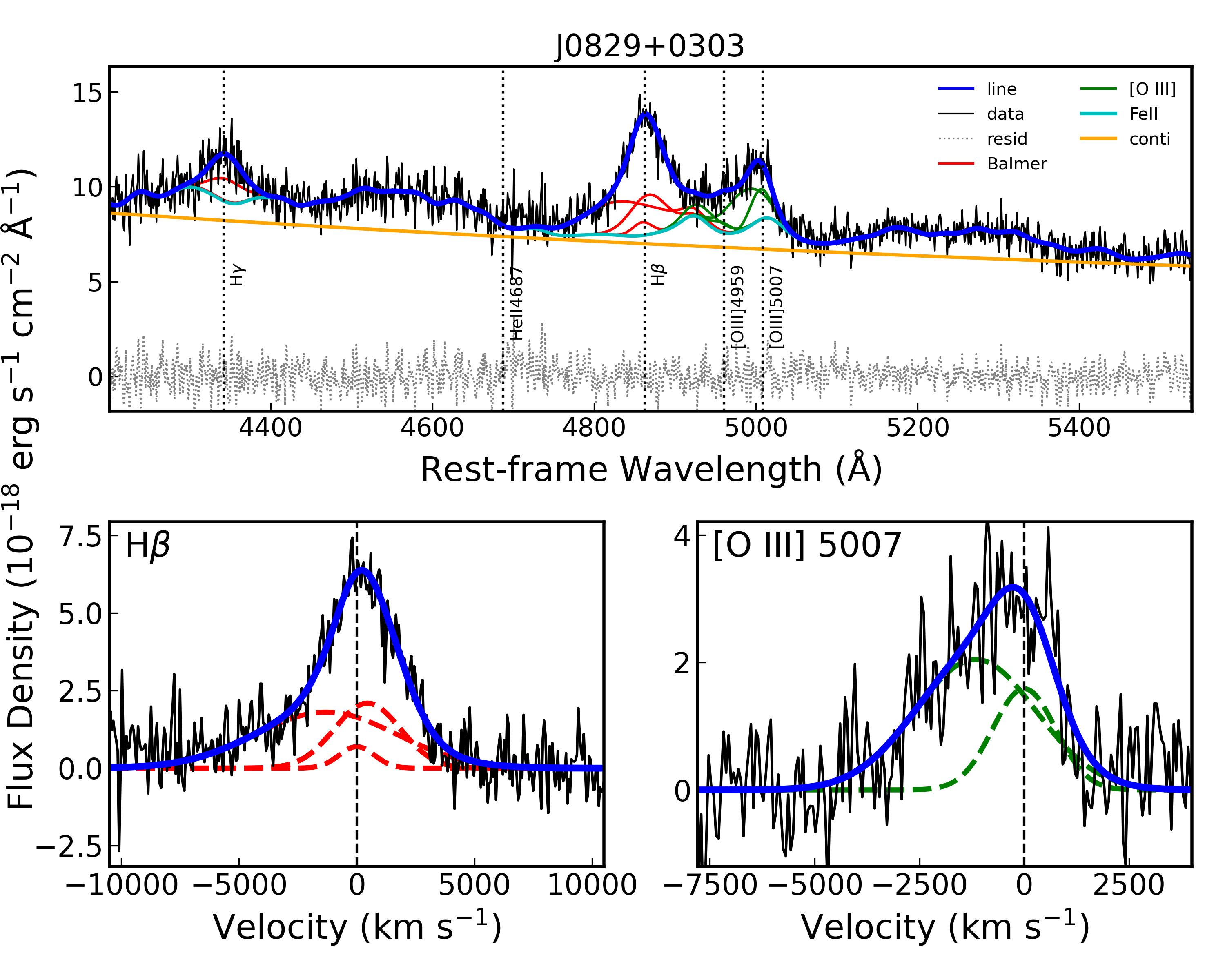}
\end{minipage}

\begin{minipage}[t]{0.5\textwidth}
\centering
\includegraphics[width=\textwidth]{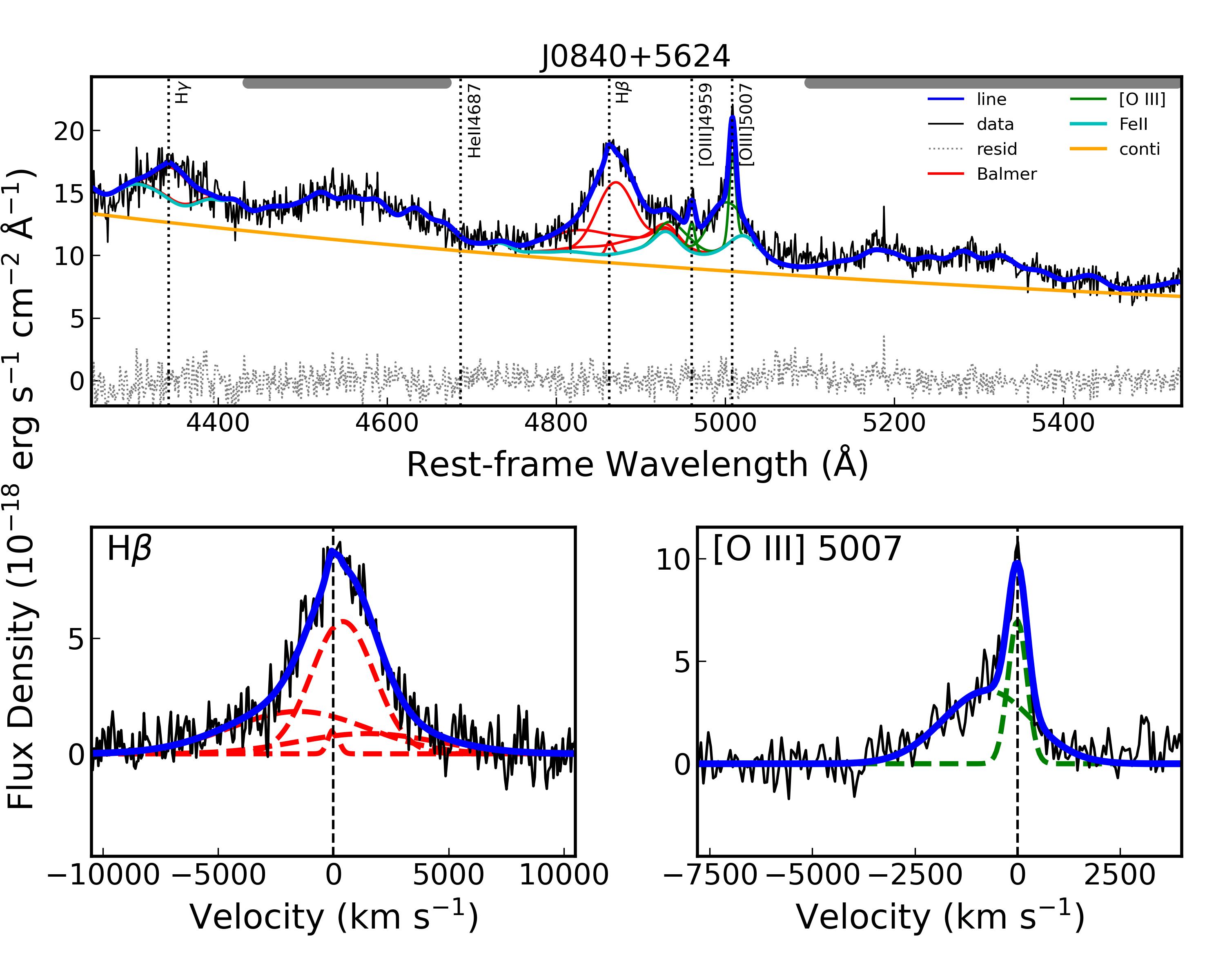}
\end{minipage}
 \begin{minipage}[t]{0.5\textwidth}
 \centering
\includegraphics[width=\textwidth]{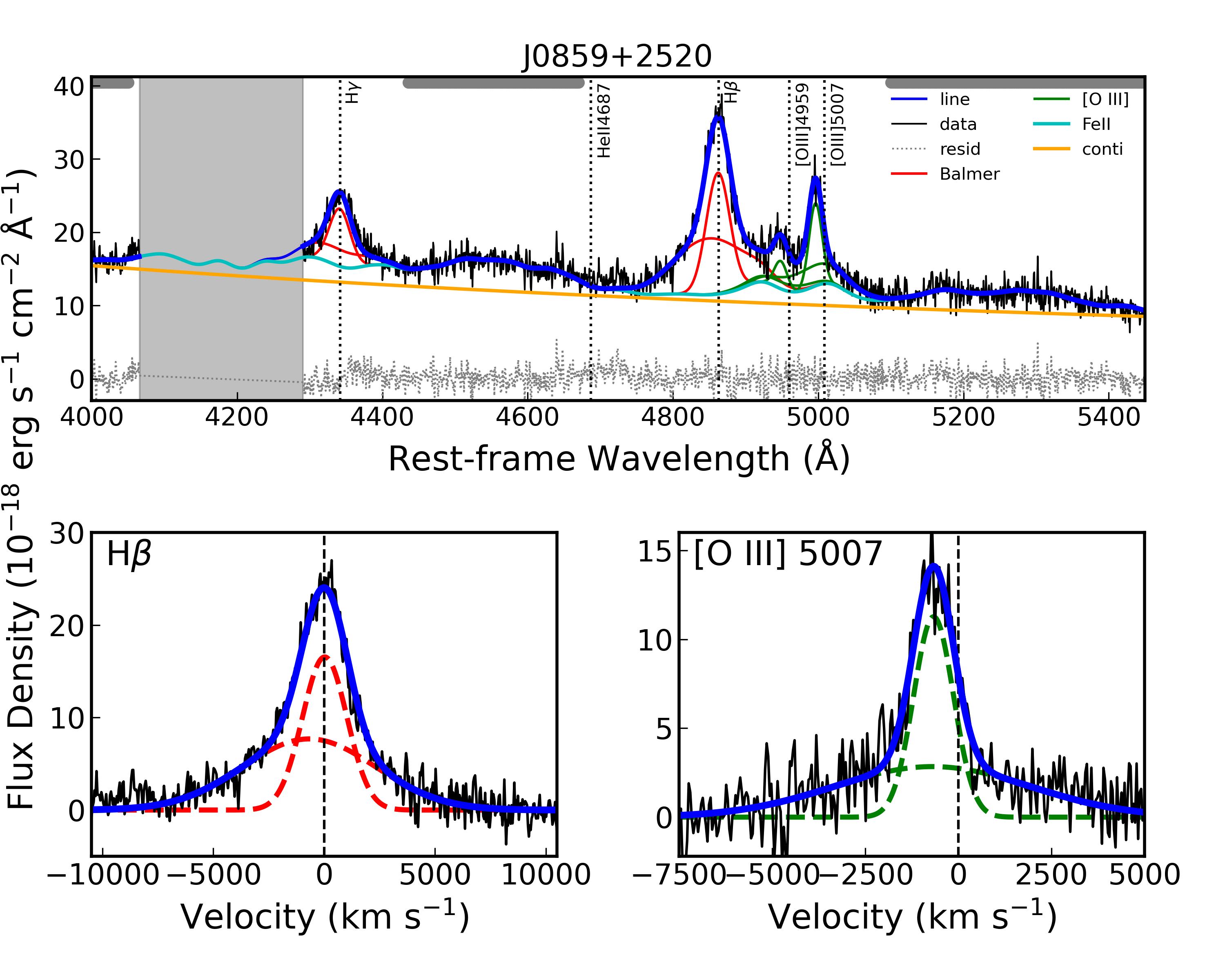}
\end{minipage}

\centering
\begin{minipage}[t]{0.5\textwidth}
\includegraphics[width=\textwidth]{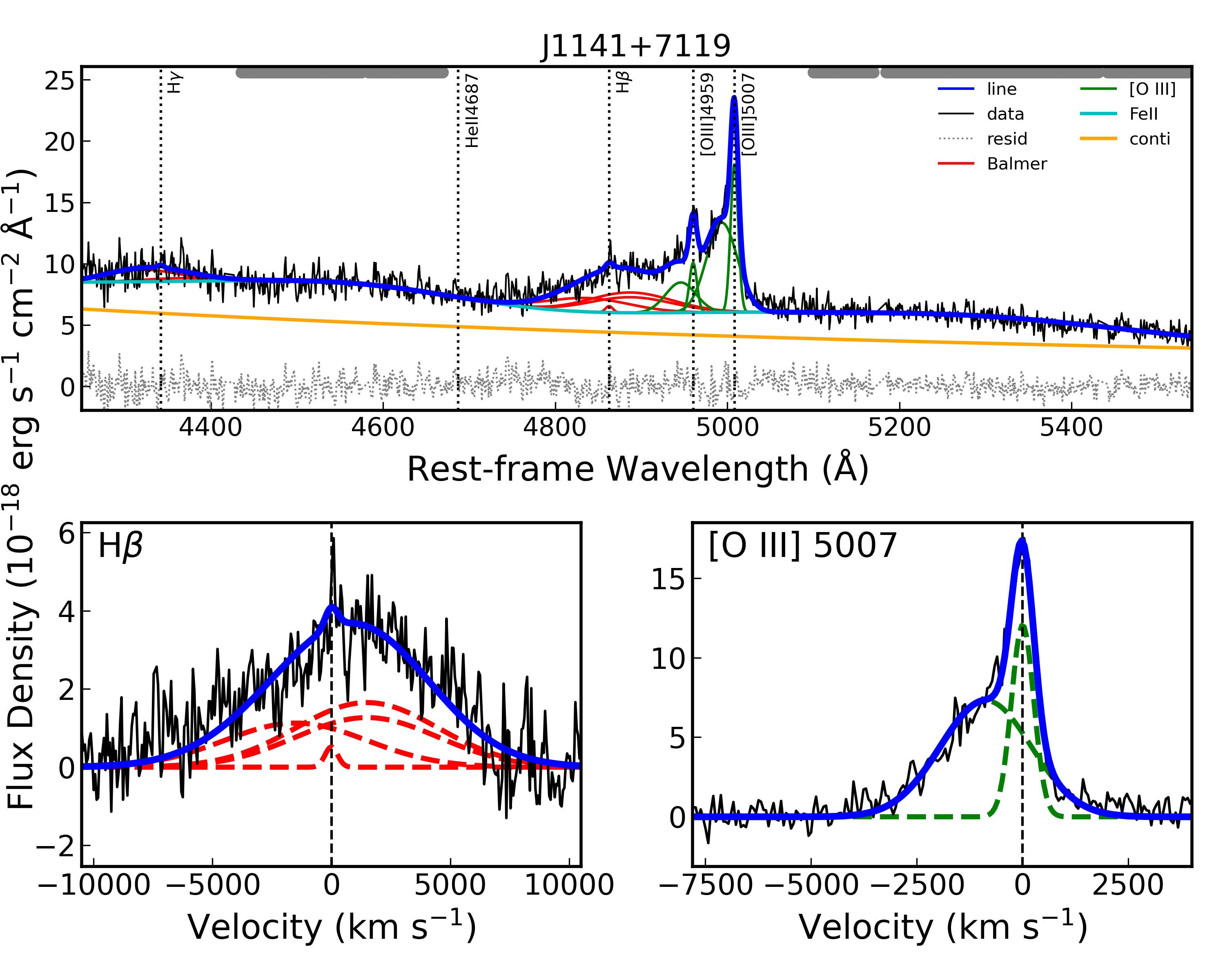}
\end{minipage}    
\caption{\textbf{Spectra of other extreme outflows.} Same as Fig. \ref{fig:profile} but for the remaining 5 extreme outflows discovered in our sample.}
\label{fig:samples}
\end{figure}

\begin{sidewaystable}
\caption{Quasar and Outflow Properties}\label{tab:outflow}%
\begin{tabular}{cll ccc cccccc}
\toprule
Object & RA & DEC & $z$ & {log($L_{\mathrm{\rm{[O~III]}}}$)} & {log($L_{\mathrm{bol}}$)} &
   \vjiuba  & 
  \wjiu & \vwu &
  {log($\dot{m}_{\mathrm{out}}$)}  &
  {log($\dot{p}_{\mathrm{out}}$)}  &
  {log($\dot{E}_{\mathrm{out}}$)} \\
 & & & & {[erg s$^{-1}$]} & {[erg s$^{-1}$]}  & {[\kms]}  & {[\kms]} & {[\kms]} &  {[\msunyr]} & {[dyne]} & {[erg s$^{-1}$]} \\
(1) & (2) & (3) & (4) & (5) & (6) & (7) & (8) & (9) & (10) & (11) & (12)
 \\\midrule
J0732+3256 & 07:32:31.28 & +32:56:18.33 & 4.771 & - & 47.5 & - & - & - & - & - & - \\
J0756+0218 & 07:56:22.37 & +02:18:20.17 & 5.762 & - & 46.8 & - & - & - & - & - & - \\
J0759+1800 & 07:59:07.57 & +18:00:54.7 & 4.796 & 44.7 & 47.2 & -6978$\pm{484}$ & 7973$\pm{421}$ & -1212$\pm{132}$ & 4.4 & 39.1 & 47.6 \\
J0807+1328 & 08:07:15.11 & +13:28:05.1 & 4.880 & 44.0 & 47.2 & -1340$\pm{48}$ & 1644$\pm{50}$ & -120$\pm{12}$ & 3.0 & 36.9 & 44.8 \\
J0829+0303 & 08:29:07.62 & +03:03:56.56 & 5.855 & 43.8 & 47.1 & -3940$\pm{326}$ & 4509$\pm{256}$ & -709$\pm{194}$ & 3.3 & 37.7 & 46.0 \\
J0831+4046 & 08:31:22.57 & +40:46:23.3 & 4.900 & - & 47.2 & - & - & - & - & - & - \\
J0840+5624 & 08:40:35.09 & +56:24:19.90 & 5.837 & 43.9 & 47.2 & -2817$\pm{92}$ & 3180$\pm{75}$ & -306$\pm{20}$ & 3.3 & 37.6 & 45.7 \\
J0850+3246 & 08:50:48.25 & +32:46:47.90 & 5.830 & - & 47.2 & - & - & - & - & - & - \\
J0859+2520 & 08:59:31.29 & +25:20:19.5 & 4.779 & 44.3 & 47.0 & -5432$\pm{252}$ & 7048$\pm{254}$ & -690$\pm{33}$ & 3.9 & 38.5 & 46.9 \\
J0927+2001 & 09:27:21.82 & +20:01:23.70 & 5.768 & - & 47.0 & - & - & - & - & - & - \\
J0941+5947 & 09:41:08.35 & +59:47:25.7 & 4.860 & 43.9 & 47.2 & -2524$\pm{8}$ & 2425$\pm{74}$ & -1010$\pm{105}$ & 3.2 & 37.4 & 45.5 \\
J0953+6910 & 09:53:55.90 & +69:10:52.62 & 5.918 & - & 47.0 & - & - & - & - & - & - \\
J1050+4627 & 10:50:05.11 & +46:27:35.5 & 4.837 & 44.2 & 47.1 & -861$\pm{11}$ & 1125$\pm{13}$ & -94$\pm{6}$ & 3.1 & 36.8 & 44.5 \\
J1100+5800 & 11:00:41.94 & +58:00:01.3 & 4.759 & 44.3 & 47.2 & -1685$\pm{22}$ & 2012$\pm{22}$ & -213$\pm{8}$ & 3.5 & 37.5 & 45.4 \\
J1102+6635 & 11:02:47.29 & +66:35:19.6 & 4.787 & - & 46.9 & - & - & - & - & - & - \\
J1116+5853 & 11:16:33.75 & +58:53:22.04 & 5.721 & 43.1 & 46.8 & -1346$\pm{33}$ & 1760$\pm{7}$ & -246$\pm{132}$ & 2.1 & 36.1 & 43.9 \\
J1134+3928 & 11:34:15.21 & +39:28:26.0 & 4.821 & - & 46.8 & - & - & - & - & - & - \\
J1141+7119 & 11:41:43.06 & +71:19:25.07 & 5.851 & 44.3 & 46.8 & -2942$\pm{25}$ & 3168$\pm{23}$ & -413$\pm{7}$ & 3.6 & 37.9 & 46.1 \\
J1245+4348 & 12:45:10.13 & +43:48:37.9 & 4.890 & - & 47.3 & - & - & - & - & - & - \\
J1257+6349 & 12:57:57.47 & +63:49:37.20 & 6.012 & - & 47.1 & - & - & - & - & - & - \\
J1327+5732 & 13:27:41.33 & +57:32:38.43 & 5.751 & 44.1 & 47.0 & -1790$\pm{52}$ & 2096$\pm{43}$ & -104$\pm{12}$ & 3.3 & 37.3 & 45.3 \\
J1328+4445 & 13:28:25.16 & +44:45:00.2 & 4.820 & 44.4 & 47.2 & -1729$\pm{61}$ & 2278$\pm{66}$ & -157$\pm{19}$ & 3.6 & 37.6 & 45.5 \\
J1342+5838 & 13:42:43.46 & +58:38:50.0 & 4.852 & 43.4 & 46.9 & -1282$\pm{46}$ & 1803$\pm{10}$ & -157$\pm{185}$ & 2.5 & 36.4 & 44.2 \\
J1436+5007 & 14:36:11.74 & +50:07:06.90 & 5.840 & 43.3 & 47.0 & -2362$\pm{14}$ & 3590$\pm{18}$ & -121$\pm{179}$ & 2.6 & 36.8 & 44.9 \\
J1458+3327 & 14:58:05.99 & +33:27:23.0 & 4.851 & 43.5 & 47.1 & -1936$\pm{34}$ & 2950$\pm{23}$ & -94$\pm{270}$ & 2.7 & 36.8 & 44.8 \\
J1620+5202 & 16:20:45.64 & +52:02:46.65 & 4.791 & 44.8 & 47.6 & -8384$\pm{161}$ & 9155$\pm{126}$ & -1521$\pm{56}$ & 4.7 & 39.4 & 48.0 \\
J1621+5155 & 16:21:00.92 & +51:55:48.79 & 5.614 & - & 47.7 & - & - & - & - & - & - \\
\bottomrule
\botrule
\end{tabular}
\footnotetext{(4): redshift; (5): \oiii\ luminosity; (6): quasar bolometric luminosity; (7) \oiiitext\ velocity where 98\% of the line flux are redshifted with respect to it; (8): \oiiitext\ line width encloses 90 percent flux; (9) \oiiitext\ velocity at the 50th percentile flux; (10): mass outflow rate; (11): momentum outflow rate. (12): kinetic energy outflow rate.} 
\end{sidewaystable}

\clearpage

\begin{table}[!htb]
\centering
\caption{\textbf{Comparison of quasar samples.} Median values with their 90\% confidence level ranges (within parentheses) and standard deviations (std.) of the quasar properties for our sample and the two comparison samples}
\begin{tabular}{cc|ccc}
\hline
 & & Our Sample & \textit{z $\sim$ 1.5--3.5} & \textit{z $<$ 1} \\
\hline
\multirow{2}{*}{log(\Lbol/erg s$^{-1}$)} & median & 47.1 (47.0--47.2) & 47.0 (47.0--47.1) & 46.8 (46.8--46.9) \\
& std. & 0.2 & 0.2 & 0.2 \\
\multirow{2}{*}{log(\mbh/\msun)} & median & 9.1 (8.9--9.3) & 9.3 (9.1--9.4) & 9.1 (9.0--9.2) \\
& std. & 0.3 & 0.3 & 0.2 \\
\multirow{2}{*}{\edd} & median & 0.9 (0.5--1.3) & 0.5 (0.3--0.6) & 0.4 (0.4--0.6)  \\
 & std. & 0.5 & 0.5 & 0.3 \\
\hline
\end{tabular}
\label{tab:comparison}
\end{table}

\bmhead{Acknowledgements}

W.L. thanks H. Guo, Q. Gu, L. Hao, J. Li, S. Mao and J. Wang for helpful discussions. W.L. acknowledges support from NASA through STScI grant JWST-Survey-3428. 
C.M. acknowledges support from Fondecyt Iniciacion grant 11240336 and the ANID BASAL project FB210003. Y.Z. acknowledges support from the NIRCam Science Team contract to the University of Arizona, NAS5-02015.
This work is based on observations made with the NASA/ESA/CSA James Webb Space Telescope. The data were obtained from the Mikulski Archive for Space Telescopes at the Space Telescope Science Institute, which is operated by the Association of Universities for Research in Astronomy, Inc., under NASA contract NAS 5-03127 for JWST. These observations are associated with program \#3428. 
Funding for the Sloan Digital Sky Survey IV has been provided by the Alfred P. Sloan Foundation, the U.S. Department of Energy Office of Science, and the Participating Institutions. SDSS-IV acknowledges support and resources from the Center for High-Performance Computing at the University of Utah. The SDSS web site is www.sdss.org.

\section*{Declarations}

\begin{itemize}
\item Conflict of interest/Competing interests 

The authors declare no competing interests.
\item Data availability 

The JWST data are from program \#3428 and available on the Mikulski Archive for Space Telescopes (MAST) at the Space Telescope Science Institute, which can be accessed via \url{http://dx.doi.org/10.17909/4t0h-0780}. The reduced 1D spectra are available at \url{https://github.com/oscarlwz/JWST_IFU_datacube/tree/main/all1dtxt_27sources}. The data for the z $\sim$ 1.5--3.5 and z $<$ 1 comparison sample can be retrieved from \url{http://quasar.astro.illinois.edu/paper_data/nirspec/} and \url{http://quasar.astro.illinois.edu/paper_data/DR16Q/}, respectively. The results for ERQs can be retrieved from \cite{Perrotta2019}.

\item Code availability 

\textit{PyQSOFit} is publicly available at \url{https://github.com/legolason/PyQSOFit}.
All other custom scripts are available at \url{https://github.com/oscarlwz/JWST_IFU_datacube.git}.

\item Author contribution

W.L. conceived the project, carried out the data reduction and analysis, and wrote the manuscript. He is also the PI of the JWST survey program \#3428 where the data come from. X.F. helped revise the manuscript and provided suggestions for data analysis and interpretation. X.F, H.L., and R.G. held one-on-one discussions with W.L. on key issues brought up by the referees. J.Y., X.J., J.L., M.P., Y.Z., E.B., S.B., T.C., T.C., R.D., A.E., H.J., M.M., C.M., J.S, Y.S., S.V., J.W., H.Z., M.Z., S.Z., M.L. reviewed and commented on the manuscript.

\item Corresponding author

Correspondence and requests for materials should be addressed to Weizhe Liu.

\end{itemize}



\begin{thebibliography}{10}
\expandafter\ifx\csname url\endcsname\relax
  \def\url#1{\burl{#1}}\fi
\expandafter\ifx\csname urlprefix\endcsname\relax\def\urlprefix{URL }\fi
\providecommand{\bibinfo}[2]{#2}
\providecommand{\eprint}[2][]{\url{#2}}
\providecommand{\doi}[1]{\url{https://doi.org/#1}}
\bibcommenthead

\bibitem{Carnall2023}
\bibinfo{author}{{Carnall}, A.~C.} \emph{et~al.}
\newblock \bibinfo{title}{{A massive quiescent galaxy at redshift 4.658}}.
\newblock \emph{\bibinfo{journal}{\nat}} \textbf{\bibinfo{volume}{619}}, \bibinfo{pages}{716--719} (\bibinfo{year}{2023}).

\bibitem{Glazebrook2024}
\bibinfo{author}{{Glazebrook}, K.} \emph{et~al.}
\newblock \bibinfo{title}{{A massive galaxy that formed its stars at z {\ensuremath{\approx}} 11}}.
\newblock \emph{\bibinfo{journal}{\nat}} \textbf{\bibinfo{volume}{628}}, \bibinfo{pages}{277--281} (\bibinfo{year}{2024}).

\bibitem{deGraff2025}
\bibinfo{author}{{de Graaff}, A.} \emph{et~al.}
\newblock \bibinfo{title}{{Efficient formation of a massive quiescent galaxy at redshift 4.9}}.
\newblock \emph{\bibinfo{journal}{Nature Astronomy}} \textbf{\bibinfo{volume}{9}}, \bibinfo{pages}{280--292} (\bibinfo{year}{2025}).

\bibitem{Dubois2013b}
\bibinfo{author}{{Dubois}, Y.} \emph{et~al.}
\newblock \bibinfo{title}{{Blowing cold flows away: the impact of early AGN activity on the formation of a brightest cluster galaxy progenitor}}.
\newblock \emph{\bibinfo{journal}{\mnras}} \textbf{\bibinfo{volume}{428}}, \bibinfo{pages}{2885--2900} (\bibinfo{year}{2013}).

\bibitem{Hartley2023}
\bibinfo{author}{{Hartley}, A.~I.} \emph{et~al.}
\newblock \bibinfo{title}{{The first quiescent galaxies in TNG300}}.
\newblock \emph{\bibinfo{journal}{\mnras}} \textbf{\bibinfo{volume}{522}}, \bibinfo{pages}{3138--3144} (\bibinfo{year}{2023}).

\bibitem{Lovell2023}
\bibinfo{author}{{Lovell}, C.~C.} \emph{et~al.}
\newblock \bibinfo{title}{{First light and reionisation epoch simulations (FLARES) - VIII. The emergence of passive galaxies at z {\ensuremath{\geq}} 5}}.
\newblock \emph{\bibinfo{journal}{\mnras}} \textbf{\bibinfo{volume}{525}}, \bibinfo{pages}{5520--5539} (\bibinfo{year}{2023}).

\bibitem{Fan2023}
\bibinfo{author}{Fan, X.}, \bibinfo{author}{Bañados, E.} \& \bibinfo{author}{Simcoe, R.~A.}
\newblock \bibinfo{title}{Quasars and the {Intergalactic} {Medium} at {Cosmic} {Dawn}}.
\newblock \emph{\bibinfo{journal}{Annual Review of Astronomy and Astrophysics}} \textbf{\bibinfo{volume}{61}}, \bibinfo{pages}{373--426} (\bibinfo{year}{2023}).
\newblock 

\bibitem{Onoue2024}
\bibinfo{author}{{Onoue}, M.} \emph{et~al.}
\newblock \bibinfo{title}{{A Post-Starburst Pathway to Forming Massive Galaxies and Their Black Holes at z$>$6}}.
\newblock \emph{\bibinfo{journal}{arXiv e-prints}} \bibinfo{pages}{arXiv:2409.07113} (\bibinfo{year}{2024}).

\bibitem{Costa2018b}
\bibinfo{author}{{Costa}, T.}, \bibinfo{author}{{Rosdahl}, J.}, \bibinfo{author}{{Sijacki}, D.} \& \bibinfo{author}{{Haehnelt}, M.~G.}
\newblock \bibinfo{title}{{Quenching star formation with quasar outflows launched by trapped IR radiation}}.
\newblock \emph{\bibinfo{journal}{\mnras}} \textbf{\bibinfo{volume}{479}}, \bibinfo{pages}{2079--2111} (\bibinfo{year}{2018}).

\bibitem{Lupi2022}
\bibinfo{author}{{Lupi}, A.}, \bibinfo{author}{{Volonteri}, M.}, \bibinfo{author}{{Decarli}, R.}, \bibinfo{author}{{Bovino}, S.} \& \bibinfo{author}{{Silk}, J.}
\newblock \bibinfo{title}{{High-redshift quasars and their host galaxies - II. Multiphase gas and stellar kinematics}}.
\newblock \emph{\bibinfo{journal}{\mnras}} \textbf{\bibinfo{volume}{510}}, \bibinfo{pages}{5760--5779} (\bibinfo{year}{2022}).


\bibitem{Bischetti2022}
\bibinfo{author}{{Bischetti}, M.} \emph{et~al.}
\newblock \bibinfo{title}{{Suppression of black-hole growth by strong outflows at redshifts 5.8-6.6}}.
\newblock \emph{\bibinfo{journal}{\nat}} \textbf{\bibinfo{volume}{605}}, \bibinfo{pages}{244--247} (\bibinfo{year}{2022}).

\bibitem{Shen2019}
\bibinfo{author}{{Shen}, Y.} \emph{et~al.}
\newblock \bibinfo{title}{{Gemini GNIRS Near-infrared Spectroscopy of 50 Quasars at z {\ensuremath{\gtrsim}} 5.7}}.
\newblock \emph{\bibinfo{journal}{\apj}} \textbf{\bibinfo{volume}{873}}, \bibinfo{pages}{35} (\bibinfo{year}{2019}).

\bibitem{Yang2021}
\bibinfo{author}{{Yang}, J.} \emph{et~al.}
\newblock \bibinfo{title}{{Probing Early Supermassive Black Hole Growth and Quasar Evolution with Near-infrared Spectroscopy of 37 Reionization-era Quasars at 6.3 {\ensuremath{\leq}} z {\ensuremath{\leq}} 7.64}}.
\newblock \emph{\bibinfo{journal}{\apj}} \textbf{\bibinfo{volume}{923}}, \bibinfo{pages}{262} (\bibinfo{year}{2021}).


\bibitem{Capellupo2011}
\bibinfo{author}{{Capellupo}, D.~M.}, \bibinfo{author}{{Hamann}, F.}, \bibinfo{author}{{Shields}, J.~C.}, \bibinfo{author}{{Rodr{\'\i}guez Hidalgo}, P.} \& \bibinfo{author}{{Barlow}, T.~A.}
\newblock \bibinfo{title}{{Variability in quasar broad absorption line outflows - I. Trends in the short-term versus long-term data}}.
\newblock \emph{\bibinfo{journal}{\mnras}} \textbf{\bibinfo{volume}{413}}, \bibinfo{pages}{908--920} (\bibinfo{year}{2011}).



\bibitem{Zhu2025}
\bibinfo{author}{{Zhu}, Y.} \emph{et~al.}
\newblock \bibinfo{title}{{Nuclear Winds Drive Large-Scale Cold Gas Outflows in Quasars during the Reionization Epoch}}.
\newblock \emph{\bibinfo{journal}{arXiv e-prints}} \bibinfo{pages}{arXiv:2504.02305} (\bibinfo{year}{2025}).

\bibitem{Spilker2025}
\bibinfo{author}{{Spilker}, J.~S.} \emph{et~al.}
\newblock \bibinfo{title}{{Direct Evidence for Active Galactic Nuclei Feedback from Fast Molecular Outflows in Reionization-era Quasars}}.
\newblock \emph{\bibinfo{journal}{\apj}} \textbf{\bibinfo{volume}{982}}, \bibinfo{pages}{72} (\bibinfo{year}{2025}).

\bibitem{Maiolino2012}
\bibinfo{author}{{Maiolino}, R.} \emph{et~al.}
\newblock \bibinfo{title}{{Evidence of strong quasar feedback in the early Universe}}.
\newblock \emph{\bibinfo{journal}{\mnras}} \textbf{\bibinfo{volume}{425}}, \bibinfo{pages}{L66--L70} (\bibinfo{year}{2012}).

\bibitem{Meyer2022}
\bibinfo{author}{{Meyer}, R.~A.} \emph{et~al.}
\newblock \bibinfo{title}{{Physical Constraints on the Extended Interstellar Medium of the z = 6.42 Quasar J1148+5251: [C II]$_{158 {\ensuremath{\mu}}m}$, [N II]$_{205 {\ensuremath{\mu}}m}$, and [O I]$_{146 {\ensuremath{\mu}}m}$ Observations}}.
\newblock \emph{\bibinfo{journal}{\apj}} \textbf{\bibinfo{volume}{927}}, \bibinfo{pages}{152} (\bibinfo{year}{2022}).

\bibitem{Bischetti2019}
\bibinfo{author}{{Bischetti}, M.} \emph{et~al.}
\newblock \bibinfo{title}{{Widespread QSO-driven outflows in the early Universe}}.
\newblock \emph{\bibinfo{journal}{\aap}} \textbf{\bibinfo{volume}{630}}, \bibinfo{pages}{A59} (\bibinfo{year}{2019}).

\bibitem{Novak2020}
\bibinfo{author}{{Novak}, M.} \emph{et~al.}
\newblock \bibinfo{title}{{No Evidence for [C II] Halos or High-velocity Outflows in z {\ensuremath{\gtrsim}} 6 Quasar Host Galaxies}}.
\newblock \emph{\bibinfo{journal}{\apj}} \textbf{\bibinfo{volume}{904}}, \bibinfo{pages}{131} (\bibinfo{year}{2020}).

\bibitem{zaka16b}
\bibinfo{author}{{Zakamska}, N.~L.} \emph{et~al.}
\newblock \bibinfo{title}{{Discovery of extreme [O III] {$\lambda$}5007 {\AA} outflows in high-redshift red quasars}}.
\newblock \emph{\bibinfo{journal}{\mnras}} \textbf{\bibinfo{volume}{459}}, \bibinfo{pages}{3144--3160} (\bibinfo{year}{2016}).

\bibitem{Liu2020}
\bibinfo{author}{{Liu}, W.} \emph{et~al.}
\newblock \bibinfo{title}{{Integral Field Spectroscopy of Fast Outflows in Dwarf Galaxies with AGNs}}.
\newblock \emph{\bibinfo{journal}{\apj}} \textbf{\bibinfo{volume}{905}}, \bibinfo{pages}{166} (\bibinfo{year}{2020}).

\bibitem{VeilleuxLiu2023}
\bibinfo{author}{{Veilleux}, S.} \emph{et~al.}
\newblock \bibinfo{title}{{First Results from the JWST Early Release Science Program Q3D: The Warm Ionized Gas Outflow in z 1.6 Quasar XID 2028 and Its Impact on the Host Galaxy}}.
\newblock \emph{\bibinfo{journal}{\apj}} \textbf{\bibinfo{volume}{953}}, \bibinfo{pages}{56} (\bibinfo{year}{2023}).

\bibitem{Marshall2023}
\bibinfo{author}{{Marshall}, M.~A.} \emph{et~al.}
\newblock \bibinfo{title}{{GA-NIFS: Black hole and host galaxy properties of two z $\simeq$ 6.8 quasars from the NIRSpec IFU}}.
\newblock \emph{\bibinfo{journal}{\aap}} \textbf{\bibinfo{volume}{678}}, \bibinfo{pages}{A191} (\bibinfo{year}{2023}).

\bibitem{Yang2023b}
\bibinfo{author}{{Yang}, J.} \emph{et~al.}
\newblock \bibinfo{title}{{A SPectroscopic Survey of Biased Halos in the Reionization Era (ASPIRE): A First Look at the Rest-frame Optical Spectra of z $>$ 6.5 Quasars Using JWST}}.
\newblock \emph{\bibinfo{journal}{\apjl}} \textbf{\bibinfo{volume}{951}}, \bibinfo{pages}{L5} (\bibinfo{year}{2023}).

\bibitem{Loiacono2024}
\bibinfo{author}{{Loiacono}, F.} \emph{et~al.}
\newblock \bibinfo{title}{{A quasar-galaxy merger at z {\ensuremath{\sim}} 6.2:
  Black hole mass and quasar properties from the NIRSpec spectrum}}.
\newblock \emph{\bibinfo{journal}{\aap}} \textbf{\bibinfo{volume}{685}},
  \bibinfo{pages}{A121} (\bibinfo{year}{2024}).


\bibitem{Decarli2024}
\bibinfo{author}{{Decarli}, R.} \emph{et~al.}
\newblock \bibinfo{title}{{A quasar-galaxy merger at z {\ensuremath{\sim}} 6.2: Rapid host growth via the accretion of two massive satellite galaxies}}.
\newblock \emph{\bibinfo{journal}{\aap}} \textbf{\bibinfo{volume}{689}}, \bibinfo{pages}{A219} (\bibinfo{year}{2024}).

\bibitem{Liu2024b}
\bibinfo{author}{{Liu}, W.} \emph{et~al.}
\newblock \bibinfo{title}{{Fast Outflow in the Host Galaxy of the Luminous z = 7.5 Quasar J1007+2115}}.
\newblock \emph{\bibinfo{journal}{\apj}} \textbf{\bibinfo{volume}{976}}, \bibinfo{pages}{33} (\bibinfo{year}{2024}).

\bibitem{Lyu2025}
\bibinfo{author}{{Lyu}, J.} \emph{et~al.}
\newblock \bibinfo{title}{{Fading Light, Fierce Winds: JWST Snapshot of a Sub-Eddington Quasar at Cosmic Dawn}}.
\newblock \emph{\bibinfo{journal}{\apjl}} \textbf{\bibinfo{volume}{981}}, \bibinfo{pages}{L20} (\bibinfo{year}{2025}).

\bibitem{yue_eiger_2024}
\bibinfo{author}{{Yue}, M.} \emph{et~al.}
\newblock \bibinfo{title}{{EIGER. V. Characterizing the Host Galaxies of Luminous Quasars at z {\ensuremath{\gtrsim}} 6}}.
\newblock \emph{\bibinfo{journal}{\apj}} \textbf{\bibinfo{volume}{966}}, \bibinfo{pages}{176} (\bibinfo{year}{2024}).

\bibitem{Shen2016}
\bibinfo{author}{{Shen}, Y.}
\newblock \bibinfo{title}{{Rest-frame Optical Properties of Luminous 1.5 $<$ Z $<$ 3.5 Quasars: The H{\ensuremath{\beta}}-[O III] Region}}.
\newblock \emph{\bibinfo{journal}{\apj}} \textbf{\bibinfo{volume}{817}}, \bibinfo{pages}{55} (\bibinfo{year}{2016}).

\bibitem{Wu2022}
\bibinfo{author}{{Wu}, Q.} \& \bibinfo{author}{{Shen}, Y.}
\newblock \bibinfo{title}{{A Catalog of Quasar Properties from Sloan Digital Sky Survey Data Release 16}}.
\newblock \emph{\bibinfo{journal}{\apjs}} \textbf{\bibinfo{volume}{263}}, \bibinfo{pages}{42} (\bibinfo{year}{2022}).






\bibitem{Cameron2011}
\bibinfo{author}{{Cameron}, E.}
\newblock \bibinfo{title}{{On the Estimation of Confidence Intervals for Binomial Population Proportions in Astronomy: The Simplicity and Superiority of the Bayesian Approach}}.
\newblock \emph{\bibinfo{journal}{\pasa}} \textbf{\bibinfo{volume}{28}}, \bibinfo{pages}{128--139} (\bibinfo{year}{2011}).

\bibitem{Utest}
\bibinfo{author}{Mann, H.~B.} \& \bibinfo{author}{Whitney, D.~R.}
\newblock \bibinfo{title}{On a test of whether one of two random variables is stochastically larger than the other} \textbf{\bibinfo{volume}{18}}, \bibinfo{pages}{50--60}.
\newblock 
\newblock \bibinfo{note}{Publisher: Institute of Mathematical Statistics}.

\bibitem{Perrotta2019}
\bibinfo{author}{{Perrotta}, S.} \emph{et~al.}
\newblock \bibinfo{title}{{ERQs are the BOSS of quasar samples: the highest velocity [O III] quasar outflows}}.
\newblock \emph{\bibinfo{journal}{\mnras}} \textbf{\bibinfo{volume}{488}}, \bibinfo{pages}{4126--4148} (\bibinfo{year}{2019}).





\bibitem{Vayner2023c}
\bibinfo{author}{{Vayner}, A.} \emph{et~al.}
\newblock \bibinfo{title}{{First Results from the JWST Early Release Science Program Q3D: Powerful Quasar-driven Galactic Scale Outflow at z = 3}}.
\newblock \emph{\bibinfo{journal}{\apj}} \textbf{\bibinfo{volume}{960}}, \bibinfo{pages}{126} (\bibinfo{year}{2024}).

\bibitem{Ross12}
\bibinfo{author}{{Ross}, N.~P.} \emph{et~al.}
\newblock \bibinfo{title}{{The SDSS-III Baryon Oscillation Spectroscopic Survey: Quasar Target Selection for Data Release Nine}}.
\newblock \emph{\bibinfo{journal}{\apjs}} \textbf{\bibinfo{volume}{199}}, \bibinfo{pages}{3} (\bibinfo{year}{2012}).

\bibitem{Hamann2017}
\bibinfo{author}{{Hamann}, F.} \emph{et~al.}
\newblock \bibinfo{title}{{Extremely red quasars in BOSS}}.
\newblock \emph{\bibinfo{journal}{\mnras}} \textbf{\bibinfo{volume}{464}}, \bibinfo{pages}{3431--3463} (\bibinfo{year}{2017}).

\bibitem{Kendall}
\bibinfo{author}{{Kendall}, M.~G.}
\newblock \bibinfo{title}{{A new measure of rank correlation}}.
\newblock \emph{\bibinfo{journal}{Biometrika}} \textbf{\bibinfo{volume}{30}}, \bibinfo{pages}{81--93} (\bibinfo{year}{1938}).

\bibitem{Costa2022}
\bibinfo{author}{Costa, T.} \emph{et~al.}
\newblock \bibinfo{title}{{AGN}-driven outflows and the formation of {Ly$\alpha$} nebulae around high-z quasars}.
\newblock \emph{\bibinfo{journal}{Monthly Notices of the Royal Astronomical Society}} \textbf{\bibinfo{volume}{517}}, \bibinfo{pages}{1767--1790} (\bibinfo{year}{2022}).
\newblock 

\bibitem{farina_requiem_2019}
\bibinfo{author}{Farina, E.~P.} \emph{et~al.}
\newblock \bibinfo{title}{The {REQUIEM} {Survey}. {I}. {A} {Search} for {Extended} {Ly$\alpha$} {Nebular} {Emission} {Around} 31 z {\textgreater} 5.7 {Quasars}}.
\newblock \emph{\bibinfo{journal}{The Astrophysical Journal}} \textbf{\bibinfo{volume}{887}}, \bibinfo{pages}{196} (\bibinfo{year}{2019}).


\bibitem{Nguyen2020}
\bibinfo{author}{{Nguyen}, N.~H.} \emph{et~al.}
\newblock \bibinfo{title}{{ALMA Observations of Quasar Host Galaxies at z $\sim$ 4.8}}.
\newblock \emph{\bibinfo{journal}{\apj}} \textbf{\bibinfo{volume}{895}}, \bibinfo{pages}{74} (\bibinfo{year}{2020}).

\bibitem{decarli_alma_2018}
\bibinfo{author}{Decarli, R.} \emph{et~al.}
\newblock \bibinfo{title}{An ALMA [C II] Survey of 27 Quasars at z $>$ 5.94}.
\newblock \emph{\bibinfo{journal}{The Astrophysical Journal}} \textbf{\bibinfo{volume}{854}}, \bibinfo{pages}{97} (\bibinfo{year}{2018}).
\newblock 



\bibitem{Wang2024}
\bibinfo{author}{{Wang}, F.} \emph{et~al.}
\newblock \bibinfo{title}{{A Spatially Resolved [C II] Survey of 31 z {\ensuremath{\sim}} 7 Massive Galaxies Hosting Luminous Quasars}}.
\newblock \emph{\bibinfo{journal}{\apj}} \textbf{\bibinfo{volume}{968}}, \bibinfo{pages}{9} (\bibinfo{year}{2024}).




\bibitem{Valentino2023}
\bibinfo{author}{{Valentino}, F.} \emph{et~al.}
\newblock \bibinfo{title}{{An Atlas of Color-selected Quiescent Galaxies at z > 3 in Public JWST Fields}}.
\newblock \emph{\bibinfo{journal}{\apj}} \textbf{\bibinfo{volume}{947}}, \bibinfo{pages}{20} (\bibinfo{year}{2023}).


\bibitem{Nanayakkara2024}
\bibinfo{author}{{Nanayakkara}, T.} \emph{et~al.}
\newblock \bibinfo{title}{{A population of faint, old, and massive quiescent galaxies at 3 $<$ z $<$ 4 revealed by JWST NIRSpec Spectroscopy}}.
\newblock \emph{\bibinfo{journal}{Scientific Reports}} \textbf{\bibinfo{volume}{14}}, \bibinfo{pages}{3724} (\bibinfo{year}{2024}).

\bibitem{Ji2026}
\bibinfo{author}{{Ji}, Z.} \emph{et~al.}
\newblock \bibinfo{title}{{JADES: Rest-frame UV-to-NIR Size Evolution of Massive Quiescent Galaxies from Redshift z = 5 to z = 0.5}}.
\newblock \emph{\bibinfo{journal}{\apj}} \textbf{\bibinfo{volume}{998}}, \bibinfo{pages}{239} (\bibinfo{year}{2026}).


\bibitem{king15}
\bibinfo{author}{{King}, A.} \& \bibinfo{author}{{Pounds}, K.}
\newblock \bibinfo{title}{{Powerful Outflows and Feedback from Active Galactic Nuclei}}.
\newblock \emph{\bibinfo{journal}{\araa}} \textbf{\bibinfo{volume}{53}}, \bibinfo{pages}{115--154} (\bibinfo{year}{2015}).

\bibitem{Harrison2018}
\bibinfo{author}{{Harrison}, C.~M.} \emph{et~al.}
\newblock \bibinfo{title}{{AGN outflows and feedback twenty years on}}.
\newblock \emph{\bibinfo{journal}{Nature Astronomy}} \textbf{\bibinfo{volume}{2}}, \bibinfo{pages}{198--205} (\bibinfo{year}{2018}).


\end{thebibliography}

\begin{thebibliography}{10}
\expandafter\ifx\csname url\endcsname\relax
  \def\url#1{\burl{#1}}\fi
\expandafter\ifx\csname urlprefix\endcsname\relax\def\urlprefix{URL }\fi
\providecommand{\bibinfo}[2]{#2}
\providecommand{\eprint}[2][]{\url{#2}}
\providecommand{\doi}[1]{\url{https://doi.org/#1}}
\bibcommenthead

\bibitem{Bok2022}
\bibinfo{author}{{B{\"o}ker}, T.} \emph{et~al.}
\newblock \bibinfo{title}{{The Near-Infrared Spectrograph (NIRSpec) on the James Webb Space Telescope. III. Integral-field spectroscopy}}.
\newblock \emph{\bibinfo{journal}{\aap}} \textbf{\bibinfo{volume}{661}}, \bibinfo{pages}{A82} (\bibinfo{year}{2022}).

\bibitem{Jak2022}
\bibinfo{author}{{Jakobsen}, P.} \emph{et~al.}
\newblock \bibinfo{title}{{The Near-Infrared Spectrograph (NIRSpec) on the James Webb Space Telescope. I. Overview of the instrument and its capabilities}}.
\newblock \emph{\bibinfo{journal}{\aap}} \textbf{\bibinfo{volume}{661}}, \bibinfo{pages}{A80} (\bibinfo{year}{2022}).

\bibitem{sdssdrq12}
\bibinfo{author}{{Ross}, N.~P.} \emph{et~al.}
\newblock \bibinfo{title}{{The SDSS-III Baryon Oscillation Spectroscopic Survey: Quasar Target Selection for Data Release Nine}}.
\newblock \emph{\bibinfo{journal}{\apjs}} \textbf{\bibinfo{volume}{199}}, \bibinfo{pages}{3} (\bibinfo{year}{2012}).

\bibitem{Wang2016}
\bibinfo{author}{{Wang}, F.} \emph{et~al.}
\newblock \bibinfo{title}{{A Survey of Luminous High-redshift Quasars with SDSS and WISE. I. Target Selection and Optical Spectroscopy}}.
\newblock \emph{\bibinfo{journal}{\apj}} \textbf{\bibinfo{volume}{819}}, \bibinfo{pages}{24} (\bibinfo{year}{2016}).

\bibitem{Yang2016}
\bibinfo{author}{{Yang}, J.} \emph{et~al.}
\newblock \bibinfo{title}{{A Survey of Luminous High-redshift Quasars with SDSS and WISE. II. the Bright End of the Quasar Luminosity Function at z {\ensuremath{\approx}} 5}}.
\newblock \emph{\bibinfo{journal}{\apj}} \textbf{\bibinfo{volume}{829}}, \bibinfo{pages}{33} (\bibinfo{year}{2016}).


\bibitem{Fan2023ext}
\bibinfo{author}{Fan, X.}, \bibinfo{author}{Bañados, E.} \& \bibinfo{author}{Simcoe, R.~A.}
\newblock \bibinfo{title}{Quasars and the {Intergalactic} {Medium} at {Cosmic} {Dawn}}.
\newblock \emph{\bibinfo{journal}{Annual Review of Astronomy and Astrophysics}} \textbf{\bibinfo{volume}{61}}, \bibinfo{pages}{373--426} (\bibinfo{year}{2023}).
\newblock 

\bibitem{NSclean}
\bibinfo{author}{{Rauscher}, B.~J.}
\newblock \bibinfo{title}{{NSClean: An Algorithm for Removing Correlated Noise from JWST NIRSpec Images}}.
\newblock \emph{\bibinfo{journal}{\pasp}} \textbf{\bibinfo{volume}{136}}, \bibinfo{pages}{015001} (\bibinfo{year}{2024}).

\bibitem{Vayner2023b}
\bibinfo{author}{{Vayner}, A.} \emph{et~al.}
\newblock \bibinfo{title}{{First Results from the JWST Early Release Science Program Q3D: Ionization Cone, Clumpy Star Formation, and Shocks in a z = 3 Extremely Red Quasar Host}}.
\newblock \emph{\bibinfo{journal}{\apj}} \textbf{\bibinfo{volume}{955}}, \bibinfo{pages}{92} (\bibinfo{year}{2023}).



\bibitem{Law2023}
\bibinfo{author}{{Law}, D.~R.} \emph{et~al.}
\newblock \bibinfo{title}{{A 3D Drizzle Algorithm for JWST and Practical Application to the MIRI Medium Resolution Spectrometer}}.
\newblock \emph{\bibinfo{journal}{\aj}} \textbf{\bibinfo{volume}{166}}, \bibinfo{pages}{45} (\bibinfo{year}{2023}).

\bibitem{Richards2002}
\bibinfo{author}{{Richards}, G.~T.} \emph{et~al.}
\newblock \bibinfo{title}{{Spectroscopic Target Selection in the Sloan Digital Sky Survey: The Quasar Sample}}.
\newblock \emph{\bibinfo{journal}{\aj}} \textbf{\bibinfo{volume}{123}}, \bibinfo{pages}{2945--2975} (\bibinfo{year}{2002}).

\bibitem{DR16Q}
\bibinfo{author}{{Lyke}, B.~W.} \emph{et~al.}
\newblock \bibinfo{title}{{The Sloan Digital Sky Survey Quasar Catalog: Sixteenth Data Release}}.
\newblock \emph{\bibinfo{journal}{\apjs}} \textbf{\bibinfo{volume}{250}}, \bibinfo{pages}{8} (\bibinfo{year}{2020}).

\bibitem{panstarr}
\bibinfo{author}{{Chambers}, K.~C.} \emph{et~al.}
\newblock \bibinfo{title}{{The Pan-STARRS1 Surveys}}.
\newblock \emph{\bibinfo{journal}{arXiv e-prints}} \bibinfo{pages}{arXiv:1612.05560} (\bibinfo{year}{2016}).

\bibitem{Wu2022ext}
\bibinfo{author}{{Wu}, Q.} \& \bibinfo{author}{{Shen}, Y.}
\newblock \bibinfo{title}{{A Catalog of Quasar Properties from Sloan Digital Sky Survey Data Release 16}}.
\newblock \emph{\bibinfo{journal}{\apjs}} \textbf{\bibinfo{volume}{263}}, \bibinfo{pages}{42} (\bibinfo{year}{2022}).

\bibitem{Shen2016ext}
\bibinfo{author}{{Shen}, Y.}
\newblock \bibinfo{title}{{Rest-frame Optical Properties of Luminous 1.5 $<$ Z $<$ 3.5 Quasars: The H{\ensuremath{\beta}}-[O III] Region}}.
\newblock \emph{\bibinfo{journal}{\apj}} \textbf{\bibinfo{volume}{817}}, \bibinfo{pages}{55} (\bibinfo{year}{2016}).

\bibitem{Shen2013}
\bibinfo{author}{{Shen}, Y.}
\newblock \bibinfo{title}{{The mass of quasars}}.
\newblock \emph{\bibinfo{journal}{Bulletin of the Astronomical Society of India}} \textbf{\bibinfo{volume}{41}}, \bibinfo{pages}{61--115} (\bibinfo{year}{2013}).

\bibitem{Kendallext}
\bibinfo{author}{{Kendall}, M.~G.}
\newblock \bibinfo{title}{{A new measure of rank correlation}}.
\newblock \emph{\bibinfo{journal}{Biometrika}} \textbf{\bibinfo{volume}{30}}, \bibinfo{pages}{81--93} (\bibinfo{year}{1938}).

\bibitem{Risaliti2011}
\bibinfo{author}{{Risaliti}, G.}, \bibinfo{author}{{Salvati}, M.} \& \bibinfo{author}{{Marconi}, A.}
\newblock \bibinfo{title}{{[O III] equivalent width and orientation effects in quasars}}.
\newblock \emph{\bibinfo{journal}{\mnras}} \textbf{\bibinfo{volume}{411}}, \bibinfo{pages}{2223--2229} (\bibinfo{year}{2011}).

\bibitem{Bisogni2017}
\bibinfo{author}{{Bisogni}, S.}, \bibinfo{author}{{Marconi}, A.} \& \bibinfo{author}{{Risaliti}, G.}
\newblock \bibinfo{title}{{Orientation effects on spectral emission features of quasars}}.
\newblock \emph{\bibinfo{journal}{\mnras}} \textbf{\bibinfo{volume}{464}}, \bibinfo{pages}{385--397} (\bibinfo{year}{2017}).

\bibitem{Vietri2018}
\bibinfo{author}{{Vietri}, G.} \emph{et~al.}
\newblock \bibinfo{title}{{The WISSH quasars project. IV. Broad line region versus kiloparsec-scale winds}}.
\newblock \emph{\bibinfo{journal}{\aap}} \textbf{\bibinfo{volume}{617}}, \bibinfo{pages}{A81} (\bibinfo{year}{2018}).

\bibitem{pyqsofit}
\bibinfo{author}{{Guo}, H.}, \bibinfo{author}{{Shen}, Y.} \& \bibinfo{author}{{Wang}, S.}
\newblock \bibinfo{title}{{PyQSOFit: Python code to fit the spectrum of quasars}}.
\newblock \bibinfo{howpublished}{Astrophysics Source Code Library} (\bibinfo{year}{2018}).
\newblock \eprint{1809.008}.

\bibitem{BorosonGreen1992}
\bibinfo{author}{{Boroson}, T.~A.} \& \bibinfo{author}{{Green}, R.~F.}
\newblock \bibinfo{title}{{The Emission-Line Properties of Low-Redshift Quasi-stellar Objects}}.
\newblock \emph{\bibinfo{journal}{\apjs}} \textbf{\bibinfo{volume}{80}}, \bibinfo{pages}{109} (\bibinfo{year}{1992}).

\bibitem{Vestergaard2001}
\bibinfo{author}{{Vestergaard}, M.} \& \bibinfo{author}{{Wilkes}, B.~J.}
\newblock \bibinfo{title}{{An Empirical Ultraviolet Template for Iron Emission in Quasars as Derived from I Zwicky 1}}.
\newblock \emph{\bibinfo{journal}{\apjs}} \textbf{\bibinfo{volume}{134}}, \bibinfo{pages}{1--33} (\bibinfo{year}{2001}).




\bibitem{VC04}
\bibinfo{author}{{V{\'e}ron-Cetty}, M.~P.}, \bibinfo{author}{{Joly}, M.} \& \bibinfo{author}{{V{\'e}ron}, P.}
\newblock \bibinfo{title}{{The unusual emission line spectrum of I Zw 1}}.
\newblock \emph{\bibinfo{journal}{\aap}} \textbf{\bibinfo{volume}{417}}, \bibinfo{pages}{515--525} (\bibinfo{year}{2004}).

\bibitem{Kovacevic2010}
\bibinfo{author}{{Kova{\v{c}}evi{\'c}}, J.}, \bibinfo{author}{{Popovi{\'c}}, L.~{\v{C}}.} \& \bibinfo{author}{{Dimitrijevi{\'c}}, M.~S.}
\newblock \bibinfo{title}{{Analysis of Optical Fe II Emission in a Sample of Active Galactic Nucleus Spectra}}.
\newblock \emph{\bibinfo{journal}{\apjs}} \textbf{\bibinfo{volume}{189}}, \bibinfo{pages}{15--36} (\bibinfo{year}{2010}).

\bibitem{Osterbrock2006}
\bibinfo{author}{{Osterbrock}, D.~E.} \& \bibinfo{author}{{Ferland}, G.~J.}
\newblock \emph{\bibinfo{title}{{Astrophysics of gaseous nebulae and active galactic nuclei}}}  (\bibinfo{publisher}{Sausalito, CA: University Science Books}, \bibinfo{year}{2006}).

\bibitem{Richards2006}
\bibinfo{author}{{Richards}, G.~T.} \emph{et~al.}
\newblock \bibinfo{title}{{Spectral Energy Distributions and Multiwavelength Selection of Type 1 Quasars}}.
\newblock \emph{\bibinfo{journal}{\apjs}} \textbf{\bibinfo{volume}{166}}, \bibinfo{pages}{470--497} (\bibinfo{year}{2006}).


\bibitem{Vestergaard2006}
\bibinfo{author}{{Vestergaard}, M.} \& \bibinfo{author}{{Peterson}, B.~M.}
\newblock \bibinfo{title}{{Determining Central Black Hole Masses in Distant Active Galaxies and Quasars. II. Improved Optical and UV Scaling Relationships}}.
\newblock \emph{\bibinfo{journal}{\apj}} \textbf{\bibinfo{volume}{641}}, \bibinfo{pages}{689--709} (\bibinfo{year}{2006}).

\bibitem{Liu2025bb}
\bibinfo{author}{{Liu}, W.} \emph{et~al.}
\newblock \bibinfo{title}{{A JWST/NIRSpec Integral Field Unit Survey of Luminous Quasars at z \raisebox{-0.5ex}\textasciitilde 5-6 (Q-IFU): Rest-frame Optical Nuclear Properties and Extended Nebulae}}.
\newblock \emph{\bibinfo{journal}{arXiv e-prints}} \bibinfo{pages}{arXiv:2511.06085} (\bibinfo{year}{2025}).

\bibitem{NFW}
\bibinfo{author}{{Navarro}, J.~F.}, \bibinfo{author}{{Frenk}, C.~S.} \& \bibinfo{author}{{White}, S. D.~M.}
\newblock \bibinfo{title}{{The Structure of Cold Dark Matter Halos}}.
\newblock \emph{\bibinfo{journal}{\apj}} \textbf{\bibinfo{volume}{462}}, \bibinfo{pages}{563} (\bibinfo{year}{1996}).

\bibitem{DuttonMaccio2014}
\bibinfo{author}{{Dutton}, A.~A.} \& \bibinfo{author}{{Macci{\`o}}, A.~V.}
\newblock \bibinfo{title}{{Cold dark matter haloes in the Planck era: evolution of structural parameters for Einasto and NFW profiles}}.
\newblock \emph{\bibinfo{journal}{\mnras}} \textbf{\bibinfo{volume}{441}}, \bibinfo{pages}{3359--3374} (\bibinfo{year}{2014}).

\bibitem{galpy}
\bibinfo{author}{{Bovy}, J.}
\newblock \bibinfo{title}{{galpy: A python Library for Galactic Dynamics}}.
\newblock \emph{\bibinfo{journal}{\apjs}} \textbf{\bibinfo{volume}{216}}, \bibinfo{pages}{29} (\bibinfo{year}{2015}).




\bibitem{Costa2024}
\bibinfo{author}{{Costa}, T.}
\newblock \bibinfo{title}{{The host dark matter haloes of the first quasars}}.
\newblock \emph{\bibinfo{journal}{\mnras}} \textbf{\bibinfo{volume}{531}}, \bibinfo{pages}{930--944} (\bibinfo{year}{2024}).

\bibitem{Wang2023}
\bibinfo{author}{{Wang}, F.} \emph{et~al.}
\newblock \bibinfo{title}{{A SPectroscopic Survey of Biased Halos in the Reionization Era (ASPIRE): JWST Reveals a Filamentary Structure around a z = 6.61 Quasar}}.
\newblock \emph{\bibinfo{journal}{\apjl}} \textbf{\bibinfo{volume}{951}}, \bibinfo{pages}{L4} (\bibinfo{year}{2023}).

\bibitem{Eilers2024}
\bibinfo{author}{{Eilers}, A.-C.} \emph{et~al.}
\newblock \bibinfo{title}{{EIGER. VI. The Correlation Function, Host Halo Mass, and Duty Cycle of Luminous Quasars at z {\ensuremath{\gtrsim}} 6}}.
\newblock \emph{\bibinfo{journal}{\apj}} \textbf{\bibinfo{volume}{974}}, \bibinfo{pages}{275} (\bibinfo{year}{2024}).

\bibitem{Wechsler2018}
\bibinfo{author}{{Wechsler}, R.~H.} \& \bibinfo{author}{{Tinker}, J.~L.}
\newblock \bibinfo{title}{{The Connection Between Galaxies and Their Dark Matter Halos}}.
\newblock \emph{\bibinfo{journal}{\araa}} \textbf{\bibinfo{volume}{56}}, \bibinfo{pages}{435--487} (\bibinfo{year}{2018}).

\bibitem{KormendyHo2013}
\bibinfo{author}{{Kormendy}, J.} \& \bibinfo{author}{{Ho}, L.~C.}
\newblock \bibinfo{title}{{Coevolution (Or Not) of Supermassive Black Holes and Host Galaxies}}.
\newblock \emph{\bibinfo{journal}{\araa}} \textbf{\bibinfo{volume}{51}}, \bibinfo{pages}{511--653} (\bibinfo{year}{2013}).

\bibitem{Liu2024bext}
\bibinfo{author}{{Liu}, W.} \emph{et~al.}
\newblock \bibinfo{title}{{Fast Outflow in the Host Galaxy of the Luminous z = 7.5 Quasar J1007+2115}}.
\newblock \emph{\bibinfo{journal}{\apj}} \textbf{\bibinfo{volume}{976}}, \bibinfo{pages}{33} (\bibinfo{year}{2024}).

\bibitem{zaka14}
\bibinfo{author}{{Zakamska}, N.~L.} \& \bibinfo{author}{{Greene}, J.~E.}
\newblock \bibinfo{title}{{Quasar feedback and the origin of radio emission in radio-quiet quasars}}.
\newblock \emph{\bibinfo{journal}{\mnras}} \textbf{\bibinfo{volume}{442}}, \bibinfo{pages}{784--804} (\bibinfo{year}{2014}).

\bibitem{Perrotta2019ext}
\bibinfo{author}{{Perrotta}, S.} \emph{et~al.}
\newblock \bibinfo{title}{{ERQs are the BOSS of quasar samples: the highest velocity [O III] quasar outflows}}.
\newblock \emph{\bibinfo{journal}{\mnras}} \textbf{\bibinfo{volume}{488}}, \bibinfo{pages}{4126--4148} (\bibinfo{year}{2019}).


\bibitem{Liu2025}
\bibinfo{author}{{Liu}, W.} \emph{et~al.}
\newblock \bibinfo{title}{{First Results from the JWST Early Release Science Program Q3D: The Fast Outflow in a Red Quasar at z = 0.44}}.
\newblock \emph{\bibinfo{journal}{\apj}} \textbf{\bibinfo{volume}{980}}, \bibinfo{pages}{31} (\bibinfo{year}{2025}).

\bibitem{Decarli2024ext}
\bibinfo{author}{{Decarli}, R.} \emph{et~al.}
\newblock \bibinfo{title}{{A quasar-galaxy merger at z {\ensuremath{\sim}} 6.2: Rapid host growth via the accretion of two massive satellite galaxies}}.
\newblock \emph{\bibinfo{journal}{\aap}} \textbf{\bibinfo{volume}{689}}, \bibinfo{pages}{A219} (\bibinfo{year}{2024}).


\bibitem{Rupke13a}
\bibinfo{author}{{Rupke}, D.~S.~N.} \& \bibinfo{author}{{Veilleux}, S.}
\newblock \bibinfo{title}{{The Multiphase Structure and Power Sources of Galactic Winds in Major Mergers}}.
\newblock \emph{\bibinfo{journal}{\apj}} \textbf{\bibinfo{volume}{768}}, \bibinfo{pages}{75} (\bibinfo{year}{2013}).


\bibitem{Marshall2023ext}
\bibinfo{author}{{Marshall}, M.~A.} \emph{et~al.}
\newblock \bibinfo{title}{{GA-NIFS: Black hole and host galaxy properties of two z $\simeq$ 6.8 quasars from the NIRSpec IFU}}.
\newblock \emph{\bibinfo{journal}{\aap}} \textbf{\bibinfo{volume}{678}}, \bibinfo{pages}{A191} (\bibinfo{year}{2023}).

\bibitem{Marshall2025}
\bibinfo{author}{{Marshall}, M.~A.} \emph{et~al.}
\newblock \bibinfo{title}{{JWST's PEARLS: A z=6 Quasar in a Train-Wreck Galaxy Merger System}}.
\newblock \emph{\bibinfo{journal}{arXiv e-prints}} \bibinfo{pages}{arXiv:2502.20550} (\bibinfo{year}{2025}).

\bibitem{Vayner2024}
\bibinfo{author}{{Vayner}, A.} \emph{et~al.}
\newblock \bibinfo{title}{{Powerful nuclear outflows and circumgalactic medium shocks driven by the most luminous quasar in the Universe}}.
\newblock \emph{\bibinfo{journal}{arXiv e-prints}} \bibinfo{pages}{arXiv:2412.02862} (\bibinfo{year}{2024}).


\bibitem{Bischetti2017}
\bibinfo{author}{{Bischetti}, M.} \emph{et~al.}
\newblock \bibinfo{title}{{The WISSH quasars project. I. Powerful ionised outflows in hyper-luminous quasars}}.
\newblock \emph{\bibinfo{journal}{\aap}} \textbf{\bibinfo{volume}{598}}, \bibinfo{pages}{A122} (\bibinfo{year}{2017}).




\bibitem{Mingozzi2019}
\bibinfo{author}{{Mingozzi}, M.} \emph{et~al.}
\newblock \bibinfo{title}{{The MAGNUM survey: different gas properties in the outflowing and disc components in nearby active galaxies with MUSE}}.
\newblock \emph{\bibinfo{journal}{\aap}} \textbf{\bibinfo{volume}{622}}, \bibinfo{pages}{A146} (\bibinfo{year}{2019}).

\bibitem{Liu2013b}
\bibinfo{author}{{Liu}, G.}, \bibinfo{author}{{Zakamska}, N.~L.}, \bibinfo{author}{{Greene}, J.~E.}, \bibinfo{author}{{Nesvadba}, N. P.~H.} \& \bibinfo{author}{{Liu}, X.}
\newblock \bibinfo{title}{{Observations of feedback from radio-quiet quasars - II. Kinematics of ionized gas nebulae}}.
\newblock \emph{\bibinfo{journal}{\mnras}} \textbf{\bibinfo{volume}{436}}, \bibinfo{pages}{2576--2597} (\bibinfo{year}{2013}).

\bibitem{Harrison2014}
\bibinfo{author}{{Harrison}, C.~M.}, \bibinfo{author}{{Alexander}, D.~M.}, \bibinfo{author}{{Mullaney}, J.~R.} \& \bibinfo{author}{{Swinbank}, A.~M.}
\newblock \bibinfo{title}{{Kiloparsec-scale outflows are prevalent among luminous AGN: outflows and feedback in the context of the overall AGN population}}.
\newblock \emph{\bibinfo{journal}{\mnras}} \textbf{\bibinfo{volume}{441}}, \bibinfo{pages}{3306--3347} (\bibinfo{year}{2014}).

\bibitem{Nakajima2023}
\bibinfo{author}{{Nakajima}, K.} \emph{et~al.}
\newblock \bibinfo{title}{{JWST Census for the Mass-Metallicity Star Formation Relations at z = 4-10 with Self-consistent Flux Calibration and Proper Metallicity Calibrators}}.
\newblock \emph{\bibinfo{journal}{\apjs}} \textbf{\bibinfo{volume}{269}}, \bibinfo{pages}{33} (\bibinfo{year}{2023}).

\bibitem{yue_eiger_2024ext}
\bibinfo{author}{{Yue}, M.} \emph{et~al.}
\newblock \bibinfo{title}{{EIGER. V. Characterizing the Host Galaxies of Luminous Quasars at z {\ensuremath{\gtrsim}} 6}}.
\newblock \emph{\bibinfo{journal}{\apj}} \textbf{\bibinfo{volume}{966}}, \bibinfo{pages}{176} (\bibinfo{year}{2024}).

\end{thebibliography}
\end{document}